%% file: llsz3.tex
\documentclass[iop]{emulateapj}

\newcommand{\noprint}[1]{}

\newcommand{\nsamp}{105}

\shorttitle{Optically-thick hydrogen at $2.6\lesssim z \lesssim 3$}
\shortauthors{Fumagalli et al.}

\begin{document}

\title{Dissecting the properties of optically-thick hydrogen at the peak of cosmic star formation history}
\thanks{This paper includes data gathered with the 6.5 meter Magellan Telescopes located at Las Campanas Observatory, Chile
and with the 3-m Shane Telescope at Lick Observatory.}

\author{Michele Fumagalli\altaffilmark{1,2,6}}
\author{John M. O'Meara\altaffilmark{3}}
\author{J. Xavier Prochaska\altaffilmark{4}}
\author{Gabor Worseck\altaffilmark{5}}

\altaffiltext{1}{Carnegie Observatories, 813 Santa Barbara Street, Pasadena, CA 91101, USA. 
  \email{mfumagalli@obs.carnegiescience.edu}}
\altaffiltext{2}{Department of Astrophysics, Princeton University, Princeton, NJ 08544-1001, USA.}
\altaffiltext{3}{Department of Chemistry and Physics, Saint Michael's College, Colchester, VT, USA.}
\altaffiltext{4}{University of California Observatories-Lick Observatory, University of California, 
  Santa Cruz, CA, USA.}
\altaffiltext{5}{Max Planck Institute for Astronomy, K{\"o}nigstuhl 17, D-69117, Heidelberg, Germany.}

\altaffiltext{6}{Hubble Fellow}

\begin{abstract}
We present results of a blind survey of Lyman limit systems (LLSs) detected in absorption against 
\nsamp\ quasars at $z\sim 3$ using the blue sensitive MagE spectrograph at the Magellan Clay telescope. 
By searching for Lyman limit absorption in the wavelength range $\lambda\sim 3000-4000$ \AA, we measure 
the number of LLSs per unit redshift $\ell(z)= 1.21 \pm 0.28$ at $z\sim 2.8$. Using a stacking analysis, 
we further estimate the mean free path of ionizing photons in the $z\sim 3$ Universe 
$\lambda^{912}_{\rm mfp} = 100 \pm 29 ~\rm h_{70.4}^{-1}~ Mpc$. Combined with our LLS survey, we conclude 
that systems with $\log N_{\rm HI} \ge 17.5~\rm cm^{-2}$ contribute only $\sim 40\%$ to the observed mean 
free path at these redshifts. Further, with the aid of photo-ionization modeling, we infer that a population 
of ionized and metal poor systems is likely required to reproduce the metal line strengths observed in
a composite spectrum of 20 LLSs with $\log N_{\rm HI} \sim 17.5 - 19~\rm cm^{-2}$ at $z\sim 2.6-3.0$. Finally, 
with a simple toy model, we deduce that gas in the halos of galaxies can alone account for the totality of 
LLSs at $z\lesssim 3$, but a progressively higher contribution from the intergalactic medium is required 
beyond $z\sim 3.5$. We also show how the weakly evolving number of LLSs per unit redshift at $z\lesssim 3$ 
can be modeled either by requiring that the spatial extent of the circumgalactic medium is redshift 
invariant in the last $\sim 10~\rm Gyr$ of cosmic evolution or by postulating that LLSs arise in halos 
that are rare fluctuations in the density field at each redshift. 
\end{abstract}


\section{Introduction}\label{intro}

The redshift distribution of the denser hydrogen clouds in the Universe is a relevant quantity for several 
cosmological studies, such as characterizing the UV extragalactic background radiation 
\citep[e.g.][]{mad99,fau09,haa12} or establishing the cosmic metal budget \citep[e.g.][]{pro06,bou07}. 
Empirically, these clouds can be easily identified as Lyman limit systems (LLSs) in spectra of background 
quasars \citep{tyt82}. By definition, these gas pockets have sufficient column density of neutral hydrogen 
($N_{\rm HI} \ge 10^{17.2}~\rm cm^{-2}$) to become optically thick at the Lyman continuum ($<912$~\AA) such 
that they imprint a characteristic spectral signature on the underlying continuum of background quasars.

Since the first discoveries of LLSs, a series of seminal surveys \citep{tyt82,sar89} have 
attempted to quantify $\ell(z)$, the redshift evolution of the number per unit redshift of 
these absorbers. In the past two 
decades, several efforts have been undertaken to expand the sample size of known LLSs and to 
better constrain $\ell(z)$ across an increasingly large redshift interval 
\citep[e.g.][]{lan91,sto94,stl95,son10}. However, these searches have yielded contrasting results, 
leaving substantial uncertainties on the actual redshift evolution of optically-thick hydrogen 
clouds \citep[see figure 16 of][]{pro10}. Possible reasons for this disagreement are
inconsistent definitions for the minimum column density at which LLSs are defined or, 
for some studies, the fact that LLSs are compiled from the literature, resulting in heterogeneous 
quasar samples. To overcome some of these limitations, new surveys or archival studies 
\citep{pro10,rib11,ome12} have focused in recent years on statistical samples of 
quasars in narrow redshift intervals.
Further, these data have been analyzed with attention to possible selection biases 
and with the goal of selecting only LLSs above a well-defined column density limit. 
As a result, $\ell(z)$ is now more accurately constrained between $z\sim 0.5-4.5$ for systems 
with $N_{\rm HI}\ge 10^{17.5}~\rm cm^{-2}$, the column density at which the optical depth at the Lyman 
limit exceeds $\tau \ge 2$. These studies have also provided us with a refined determination of another 
relevant cosmological quantity, the mean free path of ionizing photons $\lambda^{912}_{\rm mfp}$, that 
is now well measured between $z\sim 2-5$ \citep{pro09,ome12,wor13}. 

 Unfortunately, the difficulties of collecting large spectroscopic samples of quasars with coverage 
of the bluest wavelengths accessible from the ground ($\lambda \sim 3200-4000$\AA) have hampered a revised 
determination of $\ell(z)$ and $\lambda^{912}_{\rm mfp}$ between $z\sim 2.5-3.5$. Occurring at one of the 
most active epochs of galaxy evolution, this gap represents a limitation in our understanding of the
cosmic evolution of optically-thick hydrogen clouds, mainly because of the postulated connection between 
LLSs and the population of star-forming galaxies (see Section \ref{discussion}). To help complete 
our modern view of the cosmic evolution of LLSs, in this paper we present results from a spectroscopic 
survey of \nsamp\ quasars at $z\sim 3.0 \pm 0.2$, conducted with the blue sensitive MagE echellette 
spectrograph at the Magellan telescope. This survey was specifically designed to probe the properties 
of LLSs at $2.6 \lesssim z \lesssim 3$, offering an 
independent determination of $\ell(z)$ from previous surveys at similar redshifts 
\citep[][]{sar89,lan91}. Further, the higher spectral resolution allows us to identify, in addition 
to the Lyman break, part of the Lyman series as well as the associated metal lines that can be used 
to gain insight into the physical properties of LLSs \citep[e.g.][]{pro99a}.

The outline of this paper is the following. A description of the observations and data reduction is 
provided in Section \ref{obs}. In Section \ref{compos} and Section \ref{sec:mfp}, we present the MagE 
quasar composite spectrum at $z \sim 3$ together with the measurement of the mean free path of ionizing 
photons in the $z\sim 3$ Universe. The search of LLSs and the new determination of $\ell(z)$ at $z\sim 2.8$ 
is presented in Section \ref{llssurvey}, while Section \ref{complls} focuses on the chemical and ionization 
properties of the composite spectrum of a statistical sample of LLSs. Finally, a simple toy model to describe 
the connection between LLSs and galaxies is offered in Section \ref{discussion}, while a summary concludes 
this paper in Section \ref{summary}. Throughout this work, we assume a $\Lambda$CDM cosmology described by 
$\Omega_{\Lambda}=0.728$, $\Omega_{\rm m}=0.272$, and $H_0=70.4~\rm km~s^{-1}~Mpc^{-1}$ \citep{kom11}.

\section{Observations and Data Reduction}\label{obs}

With the aim of constraining $\ell(z)$ and $\lambda^{912}_{\rm mfp}$ between $2.6 \lesssim z \lesssim 3.0$, we 
have observed \nsamp\ quasars at $z_{\rm qso}\sim 3.0 \pm 0.2$ using the MagE echellette spectrograph 
($R\sim 4100$) at the Magellan Clay telescope \citep{mar08}. The high throughput of MagE down to 
$\lambda \sim 3100$ \AA\ ensures optimal coverage of the Lyman limit across the redshift range 
$z \sim 2.6-3.0$. The targeted quasars have been selected without any prior knowledge on the presence 
of absorbers along the line of sight. Only half of our sample (hereafter referred to as the color selected 
sample) has been drawn from the quasar spectroscopic catalogue of the Sloan Digital Sky Survey  
\citep[SDSS;][]{sch10}, while the remaining half of the targets (non-color selected) have been compiled 
from the literature to include only quasars that were first discovered either with radio or X-ray 
observations, or through slitless spectroscopy or optical variability.  These two different samples 
have been assembled to further explore and avoid a potential bias against sightlines free of strong 
intervening absorbers that has been recently discovered in the optical color selection algorithm 
adopted by SDSS  \citep{wor11}. 

Observations have been conducted between the year 2009 and the year 2012, as detailed in Table \ref{qsoobs}, 
for the most part under clear skies. The simple optical design of MagE  and our consistent choice of the 
$0.7''$ slit have ensured a high level of homogeneity in the sample, although observations have been 
collected across a four year period.  For each quasar, the slit was aligned with the paralactic 
angle to minimize the effects of atmospheric differential refraction. Throughout this survey, 
exposure times have been adjusted according to the quasar magnitudes and optimal scheduling of the 
observations. The final spectra have signal-to-noise ratios ($\rm S/N$)  ranging from $\sim 10-50$
per $0.4$\AA\ pixel as measured in a 100\AA\ window centered at 5500\AA. However, because of the 
different sensitivity and different dispersion across orders, the listed $\rm S/N$ provide only a 
simple description of the data quality. 

\begin{figure}
\centering
\includegraphics[angle=90,scale=0.34]{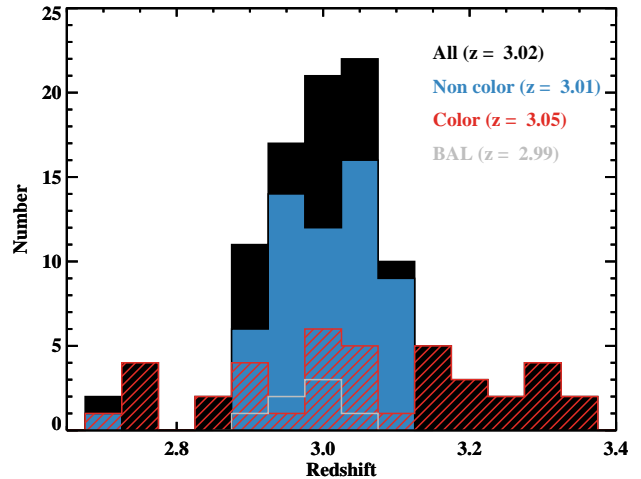}
\caption{Redshift distribution for the full quasar sample (black), and for the color and non-color selected 
samples (red and blue, respectively). The redshifts of broad absorption line quasars 
are highlighted in grey. The median of these distributions are also listed. Our survey is probing 
a narrow redshift range centered at $z_{\rm qso}\sim 3$, ideal to study the properties of optically 
thick gas between $z\sim 2.6-3.0$.}\label{reddist}
\end{figure}

Data are reduced with the {\sc mase} pipeline \citep{boc09} which, after bias subtraction and flat-fielding, 
extracts and combines the spectra in each order, producing a wavelength-calibrated spectrum in one dimension. 
Absolute flux calibration is obtained with multiple spectro-photometric stars (Feige 110, GD108, and GD50) 
that have been observed in between science exposures. Due to imperfect photometric conditions and 
slit losses, the absolute flux calibration of the 
science exposures is uncertain. However, this error does not affect our analysis which 
relies only on the relative flux calibration as a function of wavelength. A 
comparison of the fluxed spectro-photometric stars against 
template spectra reveals that the relative flux calibration of MagE data is accurate 
to within few percent between $\lambda \sim 3190-9000$ \AA. And in fact, when we compare 
colors derived from the MagE spectra for which SDSS photometry is available, we find 
excellent agreement for the $g-r$, $r-i$, $g-i$ colors (within $\sim \pm 0.03-0.05$ mag). 

Not surprisingly, a larger discrepancy ($\sim 0.3$ mag) is found for the $u-g$ color, but we exclude 
the presence of appreciable systematic errors in two ways. First, we flux calibrate multiple 
spectro-photometric stars collected during each observing night using the same sensitivity function 
that is applied to the science exposures. By comparing these fluxed spectra to the corresponding 
templates, we find agreement to within $\sim 0.1$ mag for the $u-g$ colors. Second, we compare MagE 
spectra with independent measurements made for 11 quasars\footnote{The quasars observed at 
Lick are:  HE0940-1050, SDSSJ0304-0008, SDSSJ0915+0549, SDSSJ0942+0422, 
SDSSJ0947+1421, SDSSJ1019+0825, SDSSJ1025+0452, LBQS1209+1524, LBQS1223+1753, 
Q1406+123, and TXS1033+137.} at the Shane/Lick 3 m telescope. 
During the nights February, 14-18 2013, we used the KAST spectrograph with the 
600/7500 grating, the 600/4310 grism, the D55 dichroic,  and the $2''$ slit 
under mostly clear skies. Data have been reduced following standard 
procedures\footnote{www.ucolick.org/$\sim$xavier/LowRedux/} and flux calibrated using repeated 
observations of spectro-photometric 
standard stars. Comparisons of these independent measurements reveal an excellent agreement 
in the flux calibrated spectra down to $3200$\AA. We therefore conclude that our relative 
photometric calibration is accurate to within $\sim 10-15\%$ at all wavelengths of interest to 
this study.

During our analysis, we apply a Galactic dust-extinction correction, assuming a Milky Way 
extinction curve normalized to the color excess from the dust map of \citet{sch98}. We also 
visually inspect each spectrum to identify quasars that exhibit evident broad absorption lines. 
Seven systems are found, and these are excluded from the quasar composite spectrum 
(Section \ref{compos}) and the study of the
mean free path (Section \ref{sec:mfp}). Finally, we measure quasar redshifts ($z_{\rm qso}$) by fitting 
multiple Gaussian components to the \ion{Si}{4}, \ion{C}{4}, and [\ion{C}{3}] broad emission lines. The 
resulting redshifts are listed in Table \ref{qsoobs} and are shown in Figure \ref{reddist}. The quoted 
errors and associated velocities reflect only the statistical errors and do not account for possible 
systematic discrepancies between the broad emission lines in the rest-frame UV and the rest-frame 
optical narrow emission lines. A comparison between our redshift determinations and the values listed 
in the SDSS-DR9 quasar spectroscopic catalogue \citep{par12} reveals good agreement, with a small 
median offset of $160~\rm km~s^{-1}$ which is within the measurement errors.

\begin{figure}
\centering
\includegraphics[scale=0.29]{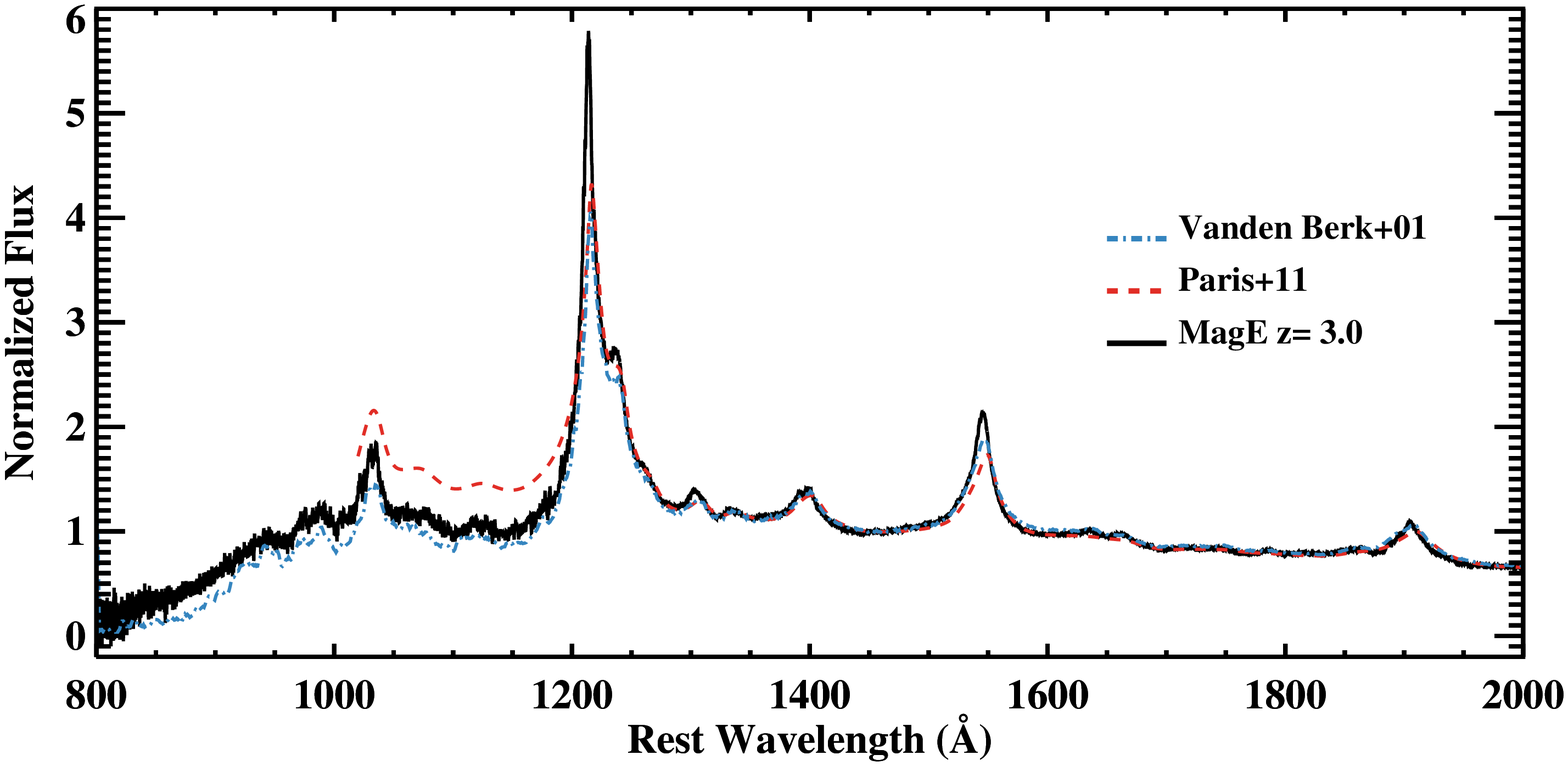}
\caption{Quasar composite spectrum of 98 MagE spectra at $z_{\rm qso} \sim 3.0 \pm 0.2$ (black solid line). 
Also shown, the $z\sim 3$ composite spectrum from \citet{par11} (red dashed line) and 
the composite spectrum by \citet{van01} (blue dotted line). There is excellent agreement in the 
continuum redward to Lyman-$\alpha$, 
while differences among the three templates are visible in the 
emission line properties. The discrepancies at $\lambda_{\rm r}<1215$~\AA\ arise instead 
from a different treatment of the IGM absorption, and from the different redshift distribution in the 
quasar samples.}\label{magestack}
\end{figure}

\section{Quasar composite spectrum}\label{compos}

Leveraging the high $\rm S/N$ and moderate resolution of these 
spectra, we construct a quasar template in a narrow redshift range ($\Delta z \sim 0.2$) centered 
around $z_{\rm qso}\sim 3$. 
Despite our small sample size, the UV sensitivity of the MagE spectrograph 
allows us to extend available composite spectra at similar redshifts \citep[e.g.][]{par11} 
down to $\sim 800$\AA\ in the quasar rest frame, without resorting to the inclusion of higher 
redshift objects as, for instance, in the \citet{van01} template. 

The final stack is produced by averaging the flux in the quasar spectra in their rest frame,
after normalizing each spectrum by the continuum value as measured in 
a 10 \AA\ window centered at 1400 \AA. The resulting composite spectrum is shown in 
Figure \ref{magestack}, together with the template by \citet{par11} and \citet{van01}.
Excellent agreement is found in the continuum reward to Lyman$-\alpha$, although each of the 
three templates exhibits different emission line properties. 
This discrepancy is expected due to variations in the quasar and host galaxy
properties (e.g. metallicity, luminosity, redshift) in the three samples, especially 
given that the MagE composite is subject to a larger sample variance than the 
templates by \citet{par11} and \citet{van01}. 
For $\lambda < 1200$\AA, instead, 
there are noticeable differences related to how the absorption from the 
intergalactic medium (IGM) has been treated. \citet{par11} correct their composite spectrum 
for the blanketing due to intervening Lyman$-\alpha$ forest clouds. 
Incidentally, we note that the ratio of the fluxes in the \citet{par11} 
template and in the MagE spectrum between $1025-1215$\AA\ is fully consistent with 
the expected IGM absorption as recently measured by \citet{bec12}.
Conversely, the flux level in the composite spectrum by \citet{van01} is 
$\sim 20\%$ lower compared to our template. We ascribe this difference to the 
inclusion of $z>3.5$ quasars combined with the rapid redshift evolution of the 
IGM transmission that changes by $\sim 25\%$ between $z\sim 3$ and $z\sim 3.5$ \citep{bec12}.

\section{The mean free path of ionizing radiation}\label{sec:mfp}

Because of the intervening IGM, the quasar composite 
spectrum at $\lambda < 1215$\AA\ differs from a simple power-law extrapolation 
of the flux redward to the Lyman$-\alpha$ emission line (Figure \ref{magestack}). 
In particular, the nearly exponential flux decrement seen at 
$<912$\AA\ contains information on the effective Lyman limit optical depth 
($\tau^{\rm LL}_{\rm eff}$) which is related to the mean free path of ionizing photons 
 ($\lambda^{912}_{\rm mfp}$). By modeling the observed composite spectrum, it 
is therefore possible to extract information on the opacity that Lyman continuum photons
experience, without the need of specifying the explicit form of the column density distribution 
function of the intervening absorbers. 

In this section, we start by reviewing the adopted formalism that was first introduced by 
\citet{pro09} and refined by \citet{ome12}, highlighting the elements that are most 
relevant to the present analysis. Next, we apply this methodology to the MagE composite 
spectrum and derive $\lambda^{912}_{\rm mfp}$ at $z\sim 3$.

\subsection{Formalism}

We start by decomposing the observed quasar flux into three components, 
\begin{equation}
F_{\rm obs}(\lambda) = A \lambda^{-\alpha} q(\lambda)  \exp(-\tau_{\rm eff})
\end{equation}
where the power law $A \lambda^{-\alpha}$ accounts for a large-scale
continuum, $q(\lambda)$ describes additional small-scale variations in the 
quasar spectral energy distribution, and $\tau_{\rm eff}$ models the wavelength-dependent 
effective optical depth arising from intervening absorbers. This last term includes the 
contribution of absorption both in the hydrogen Lyman series ($\tau^{\rm Ly}_{\rm eff}$) and, 
at $\lambda < 912$\AA, in the Lyman continuum ($\tau^{\rm LL}_{\rm eff}$)
\begin{equation}
F_{\rm obs}(\lambda) \sim A \lambda^{-\alpha} \bar{q} \exp(-\tau^{\rm LL}_{\rm eff}-\tau^{\rm Ly}_{\rm eff})\:.
\end{equation}
In this equation, we further assume that the intrinsic quasar spectral energy 
distribution $q(\lambda)$ does not evolve as a function of wavelength over the narrow 
range of interest to our study ($\sim 850-912$~\AA). Throughout this analysis, 
we further neglect additional opacity due to the presence of metal lines, which produces only 
a very modest variation in the flux \citep[$\sim 3\%$;][]{bec12}, well within the uncertainties of 
our measurement. To model the power-law continuum, we fit individual spectra in regions free from 
emission lines, between $1430-1500$\AA, $1600-1830$\AA, and $2000-2500$\AA.
The distribution of spectral indexes ($\alpha$) agrees very well with the observed 
distribution in large samples of $z_{\rm qso}\sim 3$ quasars \citep[e.g.][]{par11} with a median 
$\alpha=-1.36$.

Next, to avoid the use of an explicit form for $\tau^{\rm Ly}_{\rm eff}$ across the entire wavelength 
range, we model the continuum normalized spectrum ($\tilde{F}_{\rm obs}$) relative to the
observed value at the Lyman limit $\lambda_{912}=911.76$~\AA
\begin{equation}
f_{\rm obs}=\frac{\tilde{F}_{\rm obs}(\lambda < \lambda_{912})}{\tilde{F}_{\rm obs}(\lambda_{912})} = \frac{\exp\left(-\tau^{\rm LL}_{\rm eff}-\tau^{\rm Ly}_{\rm eff}\right)}{\exp\left(-\tau^{\rm Ly}_{\rm eff}(\lambda_{912})\right)}
\end{equation}
In this equation, we only need to provide the redshift evolution for $\tau^{\rm Ly}_{\rm eff}$, 
which in previous studies \citep[e.g.][]{ome12} has been described by a power law of index 
$\gamma_{\rm \tau}$ 
\begin{equation}\label{eqtaulya}
\frac{\tau^{\rm Ly}_{\rm eff}(\lambda)}{\tau^{\rm Ly}_{\rm eff}(\lambda_{912})}=\left(\frac{1+z_{\rm 912}}{1+z_{\rm qso}}\right)^{\gamma_{\rm \tau}}
\end{equation}
Here, $z_{912}$ is the redshift at which a photon of wavelength  $\lambda$
that is emitted at $z_{\rm qso}$ will be absorbed at the Lyman limit
\begin{equation}
z_{\rm 912} = \lambda(1 + z_{\rm qso})/\lambda_{\rm 912} - 1\:.
\end{equation}
In principle, one could fit for the free parameters in Equation (\ref{eqtaulya}), but, 
in practice, we find that data between $830-912$\AA\ do not constrain $\gamma_{\rm \tau}$
and $\tau^{\rm Ly}_{\rm eff}(\lambda_{912})$ independently. 
To account for this minor ($\sim 10-15\%$) but systematic effect, we evaluate  
$\tau^{\rm Ly}_{\rm eff}$ numerically as
\begin{eqnarray}
 \tau_{\rm eff}^{\rm Ly}(z)=\sum_{n=1}^\infty \int \int \int_z^{z_{\rm qso}} 
 f(N_{\rm HI},z',b) \times \\
\exp(-\tau_\nu^{\rm n}) dN_{\rm HI} dz' db \nonumber \:,
\end{eqnarray}
where $\tau_\nu$ is the line optical depth for the $n-$th transition in the Lyman series, 
and $f(N_{\rm HI},z)$ is the column density distribution function
from \citet{ome12} extrapolated to $z=3$. The validity of this extrapolation is 
supported by results presented in Section \ref{llssurvey}.

Once we have corrected the observed flux for the contribution of the Lyman series 
($\tilde{f}_{\rm obs}$), we only need to model the optical depth at the Lyman limit
\begin{equation}
\ln \tilde{f}_{\rm obs}=-\tau^{\rm LL}_{\rm eff}\:,
\end{equation}
which is related to the opacity $\kappa^{LL}(r,\nu)$ seen at each redshift by 
ionizing photons that have been emitted at a frequency $\nu > \nu_{912}$, 
integrated over the proper path-length $r$ from redshift $z$ to $z_{\rm qso}$
\begin{equation}
\tau_{\rm eff}^{LL}(r,\nu)=\int_0^r \kappa^{LL}(r',\nu) dr'\:.
\end{equation}
In this equation, the integrand can be rewritten as the product of the redshift-dependent opacity 
$\tilde{\kappa}_{912}(z)$ and the photo-ionization cross section $\sigma_{\rm ph} \propto \nu^{-2.75}$
which, in redshift space, becomes 
\begin{equation}
\kappa^{LL}(z) = \tilde{\kappa}_{912}(z)\left(\frac{1+z}{1+z_{912}}\right)^{-2.75}\:.
\end{equation}
For the adopted cosmology,
\begin{equation}
\frac{dr}{dz} = \frac{c/H_0}{(1+z)\sqrt{\Omega_{\rm m}(1+z)^3+\Omega_{\rm \Lambda}}}\:,
\end{equation}
where $\Omega_{\rm \Lambda}$ can be neglected between $2.7<z<3.2$ with an error of $\lesssim 3\%$. 
As previously done in the literature \citep{ome12},
we can express the evolution of $\tilde{\kappa}_{912}(z)$ with the following functional form
\begin{equation}
\tilde{\kappa}_{912}(z) = 
\tilde{\kappa}_{912}(z_{\rm qso})\left(\frac{1+z}{1+z_{\rm qso}}\right)^{\gamma_{\rm k}}\:, 
\end{equation}
but, again, we find that in practice data do not constrain $\gamma_{\rm k}$. However, because
$\tilde{\kappa}_{912}(z)$ is only weakly dependent on redshift 
\citep[$\gamma_{\rm k} \sim 0.4$ at $z\sim 2$; ][]{ome12}, 
we set $\gamma_{\rm k}=0$ for the reminder of this analysis.

In the end, the final model for the normalized flux corrected for absorption in the
 Lyman series becomes 
\begin{eqnarray}
\ln \tilde{f}_{\rm obs}= -\tau^{LL}_{\rm eff} = \frac{c~\tilde{\kappa}_{912}(z_{\rm qso})(1+z_{912})^{2.75}}{\sqrt{\Omega_{\rm m}} H_0}  \nonumber \\ \times \frac{(1+z_{\rm qso})^{-4.25}-(1+z_{\rm 912})^{-4.25}}{4.25}\:,
\end{eqnarray}
with $\tilde{\kappa}_{912}(z_{\rm qso})$ a free parameter. Finally, we define the mean free path 
$\lambda^{912}_{\rm mfp}$ as  the distance between $z_{\rm qso}$ and $z_{\rm 912}$ at 
which $\tau^{LL}_{\rm eff} = 1$.

\subsection{Results}

We apply the formalism described in the previous subsection to the MagE composite spectra of 
both the color and non-color selected quasar samples. To find the best 
$\tilde{\kappa}_{912}(z_{\rm qso})$ that describes the observed flux we adopt a 
minimum $\chi^2$ analysis in the wavelength range $830-905$\AA. The upper bound 
of this interval is 
chosen to avoid the quasar proximity region at $>905$\AA, while the 
lower bound is set to avoid noisy data ($\rm S/N \lesssim 3$ per pixel). Although we are 
selecting an optimal wavelength region, we have verified that the best fit value for 
$\lambda^{912}_{\rm mfp}$ is largely insensitive to our exact choice of the 
spectral range included in the analysis.

\begin{figure}
\centering
\includegraphics[scale=0.34,angle=90]{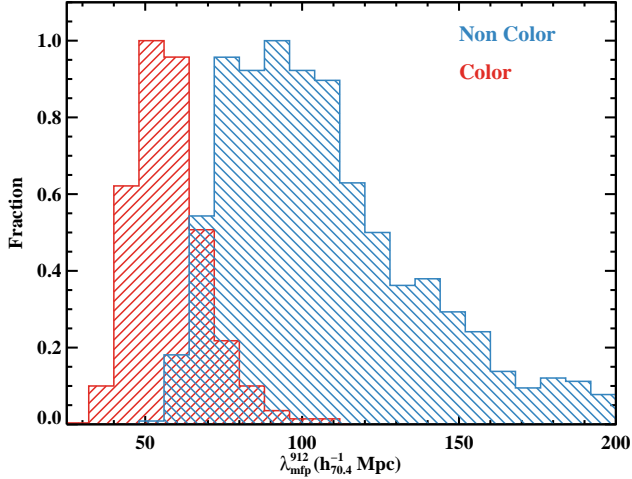}
\caption{Distribution of the best fit values for $\lambda^{912}_{\rm mfp}$ in 1000
realizations of the quasar composite spectrum, allowing for repetitions. Both results 
from the color (red) and non-color (blue) selected samples are shown. Because of a 
selection bias that systematically avoids blue quasars, the mean free path inferred 
for the color selected sample is systematically lower than what is measured in the 
composite of non-color selected quasars.}\label{mfphist}
\end{figure}

\begin{figure}
\centering
\includegraphics[angle=90,scale=0.34]{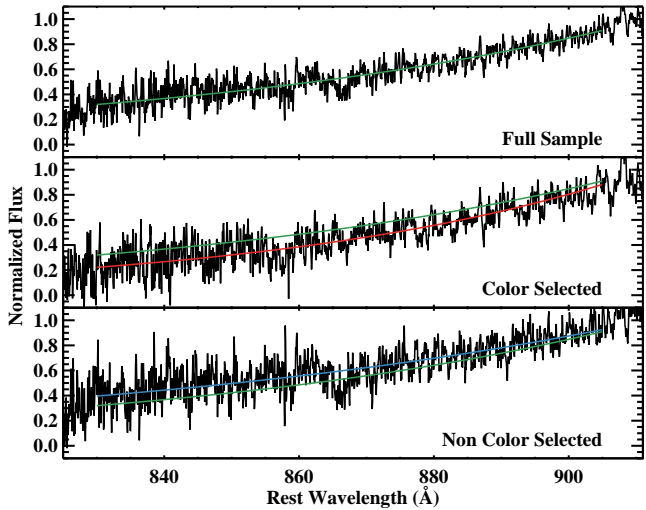}
\caption{Zoom-in of the quasar composite spectra normalized at 912\AA\ for 
the full sample (top, 98 quasars) and for the color (middle, 40 quasars) and non-color 
(bottom, 58 quasars) selected sample. The best fit models for the transmitted Lyman continuum 
flux (shown in green, red, and blue respectively) are superposed to emphasize the lower 
transmitted flux in color selected quasars compared to non-color selected ones.}\label{cfrmfpconco}
\end{figure}

Errors on $\lambda^{912}_{\rm mfp}$ are derived via bootstrapping by modeling
1000 composite spectra from the color and non-color selected samples which we construct 
as described in Section \ref{compos}, but allowing for repetitions. 
Each composite spectrum is then normalized to the observed value at $912\pm0.2$~\AA, 
after introducing a random variation up to $\pm 10\%$ in the normalization constant 
to account for additional uncertainties in the flux calibration. The two distributions of the 
best fit values for $\lambda^{912}_{\rm mfp}$ are shown in Figure \ref{mfphist}.

\begin{figure}
\centering
\includegraphics[angle=90,scale=0.34]{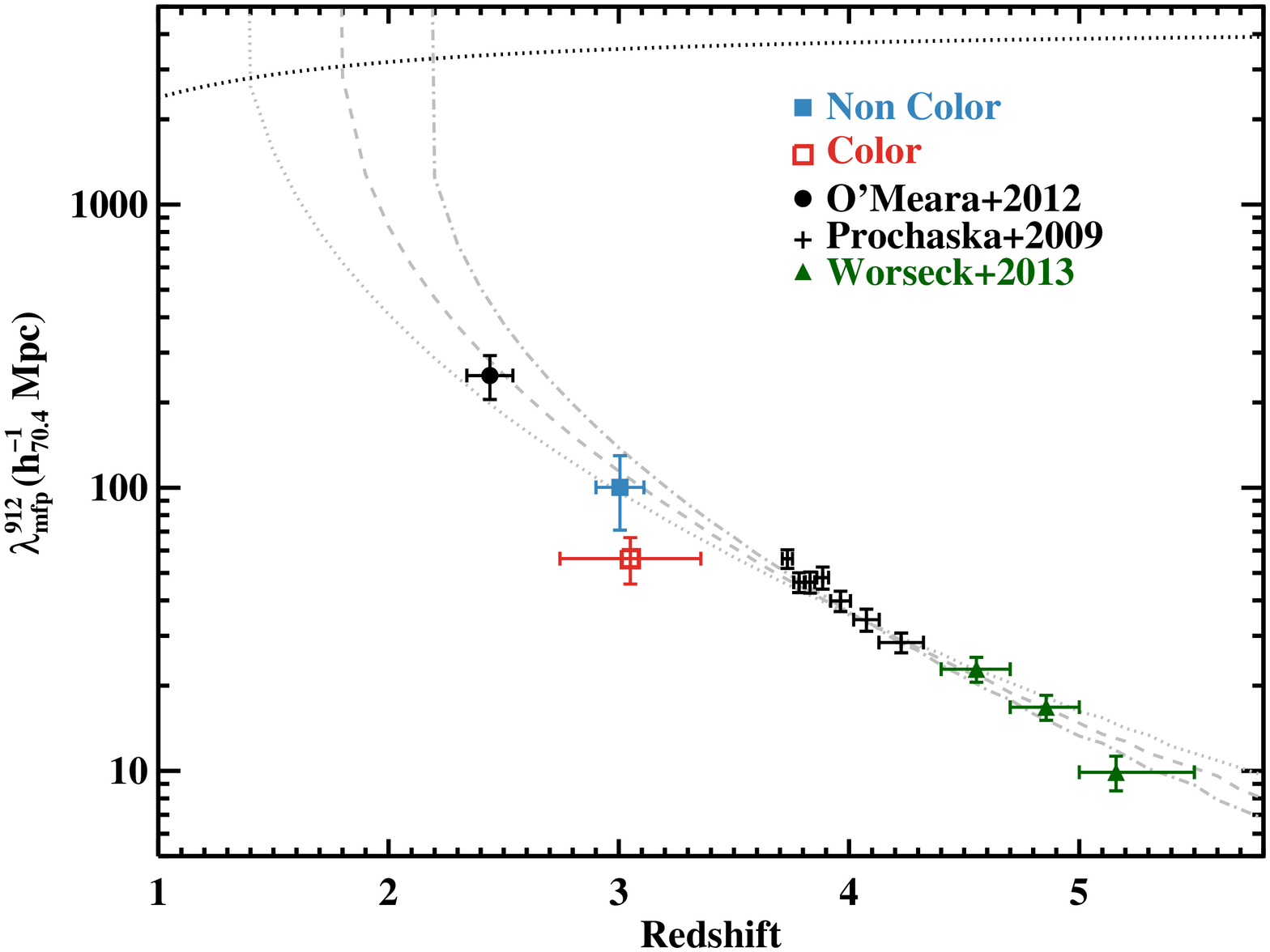}
\caption{The redshift evolution of the mean free path of ionizing photons. Values 
  measured in the color and non-color selected samples at $z\sim 3$ are shown with 
  open red and filled blue squares, respectively. Measurements at higher and lower redshifts 
  from the literature are also shown \citep{pro09,ome12,wor13}. The horizon 
  at which ionizing photons freely propagate in the Universe is marked by a black dotted line. 
  Grey lines mark instead the analytic value of $\lambda^{912}_{\rm mfp}$ derived using the 
  $f(N_{\rm HI},X)$ from \citet{ome12}. The normalization of the column density distribution 
  function is modeled as a power law of index $\gamma=2.5,2,1.5$ 
  from top to bottom. Observations favor a smooth redshift evolution with 
  index $\gamma \sim 2-1.5$, such that ionizing photons can freely propagate in the Universe by 
  $z \sim  1.4-1.8$.}\label{mfpred}
\end{figure}

\renewcommand{\thefigure}{\arabic{figure}.2} 

\begin{figure*}
\centering
\includegraphics[scale=0.4]{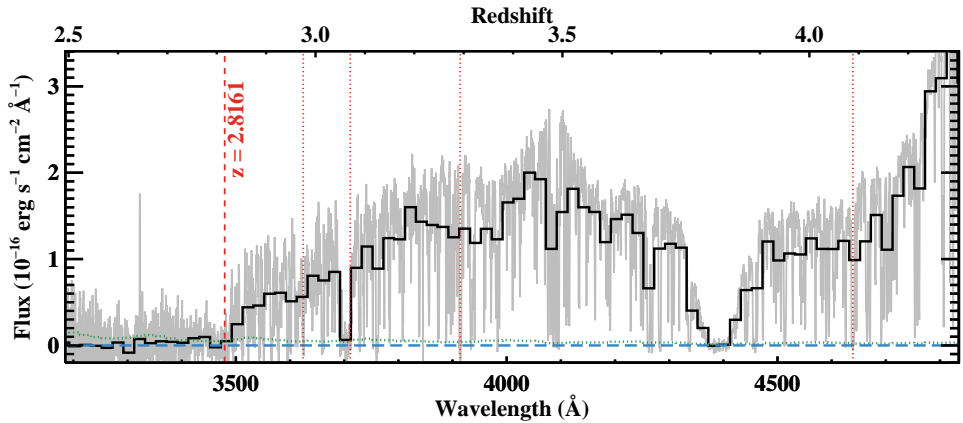}
\caption{Blue portion of the MagE spectrum of the quasar Q0038-4041. Data at full resolution 
are shown in grey. The spectrum in bins of 20\AA\ is shown in black and the green dotted line
represents the associated error array. The horizontal blue line marks the zero-flux level, 
while the red vertical lines mark the Lyman limit (dashed line) and the first four transitions 
of the Lyman series (dotted lines) of the intervening LLS. The top axis translates wavelength 
in redshift at $\lambda = 911.76$\AA. Similar figures for all the quasars are available in the electronic 
edition of the journal (Figures 6.1-6.105).}\label{egspectrum}
\end{figure*}

\renewcommand{\thefigure}{\arabic{figure}}

There is a significant difference
in the distributions of $\lambda^{912}_{\rm mfp}$ measured in color and non-color selected quasars.
A Kolmogorov-Smirnov (KS) test for the color and non-color selected samples  
returns a KS statistic $D=0.81$, and the associated probability
that the two distributions in Figure \ref{mfphist} are drawn from the same parent population 
is $\ll 10^{-5}$. This systematic discrepancy is not only evident from the two distributions of 
$\lambda^{912}_{\rm mfp}$, but it is also noticeable when inspecting the observed flux at 
$<912$\AA\ in the composite spectra (Figure \ref{cfrmfpconco}).

The different mean free path originates from the SDSS color selection algorithm that 
systematically misses blue quasars, a bias that has been previously discussed in \citet{pro09} and 
\citet{wor11}. \citet{wor11} used mock quasar photometry 
to show that the SDSS color selection  avoids 
quasars with $u-g \lesssim 2$ between $3 \lesssim z \lesssim 3.6$.
They also concluded that, because the presence of a LLS along the line of sight ``reddens'' 
the quasar photometry shifting its color far from the stellar locus, SDSS preferentially 
selects quasars with intervening absorbers. Our study of the mean free path in samples of
color and non-color selected quasars, together with the survey of LLSs in the next 
section, confirms the presence of this bias. For this reason, in the reminder of this paper 
we focus our statistical analysis on the non-color selected samples, providing results 
from the color selected sample for comparison only.

Figure \ref{mfphist} also reveals that both distributions of  $\lambda^{912}_{\rm mfp}$
exhibit a tail towards high values of $\lambda^{912}_{\rm mfp}$. We interpret this feature, 
visible in other studies \citep[e.g.][]{ome12}, as a consequence of sample variance in a 
set of quasars with a discrete number of intervening LLSs. 
As previously done in the literature, we express our best determination of the mean 
free path in the $z\sim3$ Universe using the median value of these distributions, 
together with the standard deviation computed after excluding significant outliers in the 
tail. The mean free path from both the color 
and non-color selected samples are $\lambda^{912}_{\rm mfp}= 56 \pm 10 ~\rm h^{-1}_{70.4}~\rm Mpc$ and 
$\lambda^{912}_{\rm mfp}= 100 \pm 29 ~\rm h^{-1}_{70.4}~Mpc$. 
The 5th, 25th, 75th, and 95th percentiles of the distribution of $\lambda^{912}_{\rm mfp}$ 
for the non-color selected sample are 68,  84, 126, and 184 respectively (in units of 
$h^{-1}_{70.4}~\rm Mpc$).

Our new determination is higher but statistically consistent with the mean free path estimate 
of \citet{fau08} at comparable redshift. These authors use an independent technique 
based on an analytic calculation for a given column density distribution function and find 
$\lambda^{912}_{\rm mfp} \sim 85\rm ~Mpc$ at $z\sim 3$. 
In Figure \ref{mfpred}, our new mean free path estimates  are also compared to 
the results of other studies that are based on a similar stacking analysis 
\citep{pro09,ome12,wor13}. This figure shows that the mean free path of ionizing photons 
in the Universe smoothly evolves between $z\sim 2.5-5.5$. This redshift evolution is well 
described analytically by
\begin{eqnarray}\label{tauanalyt}
\tau_{\rm eff}^{LL}(z,z_{\rm qso}) = \int^{\infty}_{0}\int^{z_{\rm qso}}_z f(N_{\rm HI},z') 
\nonumber \\ \times  \left(1-\exp\left(-N_{\rm HI}\sigma_{\rm ph}\right)\right) dN_{\rm HI} dz'\:,
\end{eqnarray}
with $\sigma_{\rm ph}$ the frequency-dependent photoionization cross section and 
$f(N_{\rm HI},z)$ the 5-parameter column density distribution function by \citet{ome12}. 
Here, $f(N_{\rm HI},z)$ has a redshift invariant shape, but the normalization evolves 
as a power law of index $\gamma \sim 1.5-2$, with $\gamma \sim 2$ slightly favored 
by the data (see grey dashed line in Figure \ref{mfpred}). According to this analytic form, 
that we normalize to match the observed $\lambda^{912}_{\rm mfp}$ at $z\sim4$, photons can freely 
escape from the emitting sources by $z\sim 1.4-1.8$, the redshift at which the mean free 
path of ionizing photons equals the horizon $h=\int_{r(0)}^{r(z_{\rm qso})} dr'$.

Before concluding our discussion on the mean free path, we note that recent work by 
\citet{rud13} suggests a significantly shorter $\lambda^{912}_{\rm mfp}$ at $z \sim 2.4$ 
compared to the estimate of \citet{ome12}. This result would imply a shallower 
redshift evolution than what is presented in Figure \ref{mfpred}.
The analysis of \citet{rud13} relies on the measurement of the column density distribution 
function at $z\sim 2.4$, for which discrepant values exist in the literature. 
Due to the difficulties of measuring individual absorption line systems 
at column density above $N_{\rm HI} \gtrsim 10^{15.5}~\rm cm^{-2}$ where saturation 
effects in the strongest lines of the Lyman series arise, we currently prefer 
the more direct measurements of the mean free path from the stacking analysis.

\section{A survey of $\tau \ge 2$ Lyman Limit systems}\label{llssurvey}

\subsection{Catalogue of intervening LLSs}

We compile a catalogue of intervening LLSs in two steps.
First, we perform an automated search for flux discontinuities that can be 
attributed to absorption at the Lyman limit in the blue portion of the MagE 
spectra. This is done by comparing the observed flux in bins of 18\AA\ to a
model for the quasar continuum that is constructed by extrapolating the power law 
continuum measured at $>1215$\AA\ and including absorption from the Lyman-$\alpha$ forest.
Starting from the Lyman limit at the quasar redshift and moving blueward, the 
automatic procedure searches for flux discontinuities that can be modeled with 
an intervening LLS with $N_{\rm HI} \ge 10^{17}~\rm cm^{-2}$ in steps of redshift 
$\Delta z = 0.01$. Once a flux break is detected, the procedure compares models 
for the Lyman limit absorption at different column densities 
(in steps of $\Delta \log N_{\rm HI} = 0.1$) to identify $\tau \ge 2$ LLSs. 
Redshifts and column densities for these systems are stored.

To obtain a clean and complete census of LLSs at $2.6 \lesssim z \lesssim 3$, we then inspect each 
quasar spectrum in search of a sharp flux decrement associated to the Lyman limit in the rest frame 
of $\tau \ge 2$ LLSs (e.g. Figure \ref{egspectrum}), starting from the 
list of systems flagged by the automated search. Three of the authors (MF, JO, XP) have independently 
modeled the observed quasar flux by superimposing to an underlying continuum level both the Lyman series 
and the Lyman limits associated to intervening absorption systems. During this procedure, two different 
methods for establishing the quasar continuum have been adopted. The first method, previously used 
by \citet{ome12}, relies on the \citet{tel02} composite spectrum, which we adapt to the 
observed flux by allowing for a scale in normalization and a power law tilt. The second method, 
instead, assumes an extrapolation of the power law continuum measured at $>1215$\AA, to which 
we apply a correction for the Lyman-$\alpha$ forest absorption as described in Section 
\ref{compos}. Both methods are found to yield consistent results.
 We emphasize that during this search, systems are selected purely based on their Lyman limit and 
Lyman series. Redshifts are determined by means of high order lines in the Lyman series to avoid 
a possible bias associated to the presence or lack of metal lines. Further, particular attention is given to 
the transmitted flux blueward of the Lyman limit which is needed to separate LLSs at 
$\tau \ge 2$ from those at $\tau < 2$.

\begin{figure}
\centering
\includegraphics[angle=90,scale=0.34]{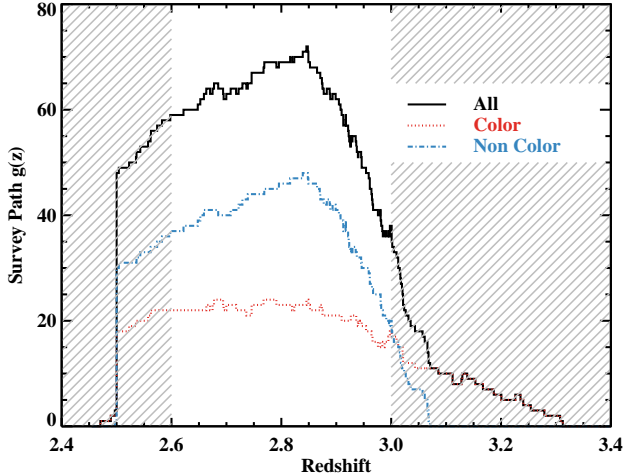}
\caption{Redshift path probed by our MagE survey of $\tau \ge 2$ LLSs for the full 
(black solid line), the color selected (red dotted line), and the non-color selected 
(blue dashed line) samples. The grey shaded area highlights the redshift range excluded 
from the statistical analysis either because of the limited path ($z > 3$)
or because of the lower $\rm S/N$ at $z<2.6$.}\label{survpath}
\end{figure}

\begin{figure}
\centering
\includegraphics[angle=90,scale=0.34]{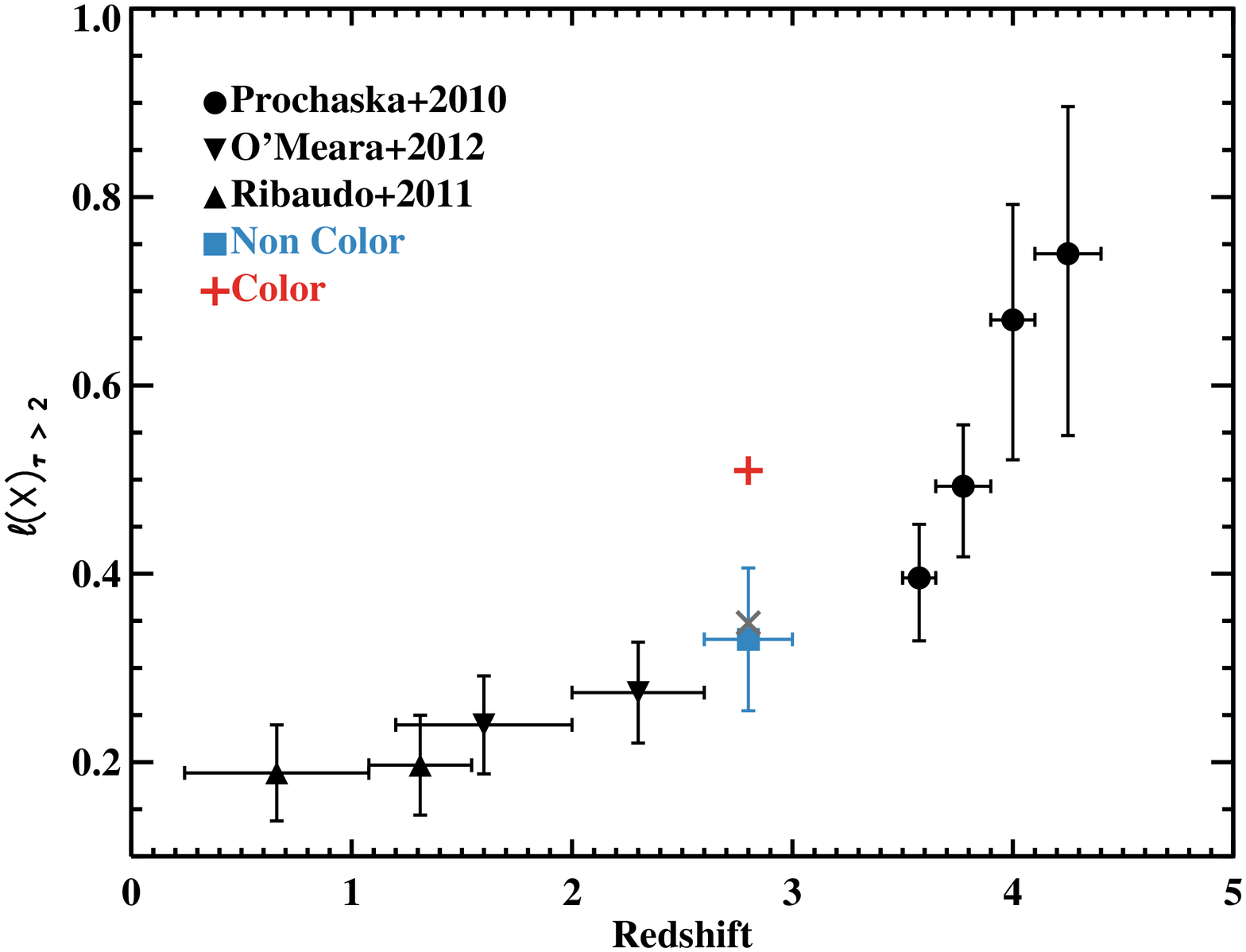}
\caption{Redshift evolution of the number of $\tau \ge 2$ LLSs per unit redshift. The result of 
  our MagE survey (blue square for the non-color selected sample) is compared to values from the 
  literature \citep{ome12,pro10,rib11} at lower and higher redshifts. $\ell(X)$ measured in the 
  color selected sample is also shown with a red cross. The grey X shows instead the expected 
  value of $\ell(X)$ computed by 
  extrapolating the \citet{ome12} $f(N_{\rm HI},X)$ to $z=2.8$. Observations 
  reveal a smooth evolution in the number density of LLSs between $z\sim 0.5-4.5$.}\label{lofxtau2}
\end{figure}

For the majority of cases ($99/105$), at least two authors have agreed on the presence and redshift of a $\tau \ge 2$ LLS. For the remaining  $6/\nsamp$ cases in which discrepant models have been produced, three authors  have jointly discussed the best model to describe the data. Not surprisingly, these discrepancies arise mostly at low $\rm S/N$ and blueward of $\lesssim 3400$\AA. We estimate that typical uncertainties of the hydrogen column density at $N_{\rm HI} = 10^{17.5}~\rm cm^{-2}$ range between $\sim 0.05-0.10$ dex, and depend on the spectrum $\rm S/N$. However, the quality of the data at $\lesssim 3300$\AA\ quickly worsen, and a much larger uncertainty is associated to systems at $z<2.6$. Because of our limited ability in identifying $\tau \ge 2$ LLSs at these redshifts, in the following analysis we restrict our statistical survey to $2.6 \le z \le 3.0$.

We further assess the completeness and purity of our catalogue of LLSs at $z \ge 2.6$
using mock spectra. One of the authors (GW) generated four independent realizations of 
MagE spectra for each of the 105 quasars, assuming the same continuum and noise properties 
of the individual observed  spectra and following the procedures detailed in \citet{dal08}, 
\citet{pro10}, and \citet{wor11}. These 420 mock spectra were subsequently analyzed independently 
by the other three authors (MF, JO, XP), following the same procedures adopted for the real data 
as described above. Results were then compared to the input catalogue of 
LLSs. From this comparison, we conclude that in $95\%$ of the cases (398/420 spectra), we are able 
to either correctly identify the presence of a $\tau \ge 2$ LLS or to establish that no 
optically-thick system is present along the line of sight. Further, 
we successfully identify 189 of the 202 $\tau \ge 2$ 
LLSs that are embedded in the mock spectra,  corresponding to 
a completeness of $94\%$. Thanks to the presence of the Lyman series, we are able to 
characterize the redshift of $90\%$ of the identified LLSs to within $\delta z = \pm 0.0005$. 
Finally, 198 of the 207 systems that we flag as $\tau \ge 2$ LLSs have in fact a column density 
$\log N_{\rm HI} \ge 17.5$, corresponding to a purity of $96\%$. The remaining 9 systems 
are false positive that have column densities $17.4 \le \log N_{\rm HI} < 17.5$, corroborating 
our estimate for the error on the column density of $\sim 0.05-0.1$ dex.

\subsection{Incidence of LLSs}

Having a complete list of LLSs in our sample (Table \ref{sellls}), we can extract a statistical sample
to quantify the number of LLSs per unit redshift by imposing additional selection criteria
that define the total redshift path $g(z)$ probed by our MagE survey (Figure \ref{survpath}). 
First, LLSs with a velocity $\le 3000~\rm km~s^{-1}$ from the quasar redshift are excluded because of 
their likely physical proximity to the quasar host galaxies. Further, when a $\tau \ge 2$ LLS is detected at 
$z_{\rm lls}$, the sharp flux decrement prevent us from searching additional systems at $z<z_{\rm lls}$. 
Therefore,  in each sightline, we consider the redshift range 
$2.6 \equiv z_{\rm end} < z < z_{\rm start} \equiv z_{\rm qso} - z_{\rm 3000 km/s}$. If a $\tau \ge 2$ LLS
is detected, $z_{\rm end}=z_{\rm lls}$, and if $z_{\rm lls}>z_{\rm start}$, $z_{\rm end}=z_{\rm start}$.
Finally, because of the limited path of our survey beyond $z=3$, we impose a limit $z_{\rm start}\le 3$.
Individual values for $z_{\rm end}$ and $z_{\rm start}$ are summarized in Table \ref{sellls}, where 
we also highlight quasars that are in common with the previous surveys by 
\citet{sar89} and \citet{lan91}. Considering the small overlap (10/105 sightlines), 
this survey is largely independent from previous efforts at comparable redshifts.

Given the total redshift path of the survey, we define the incidence of $\tau \ge 2$ LLSs as 
the total number of systems detected within the useful redshift range divided by the integral of
$g(z)$
\begin{equation}
\ell(z)_{\tau \ge 2} = \frac{N_{\rm lls}}{\int g(z) dz}\:.
\end{equation}
In our statistical sample between $2.6\le z \le 3$, we find $\ell(z)_{\tau \ge 2} = 1.21 \pm 0.28$ 
for the non-color selected sample and $\ell(z)_{\tau \ge 2} = 1.86 \pm 0.46$ for the color selected sample. 
Here, the uncertainties are related to counting errors only. As previously found for the mean free path (Figure \ref{mfphist}), 
there is a systematic difference between the color and non-color selected samples, albeit the two 
values of $\ell(z)$ are marginally consistent due to the large uncertainties. 

For the assumed cosmological model, we can further use the identity 
$\ell(z) dz = \ell(X) dX$ to recast the observed number of LLSs per unit redshift 
into a more physically motivated quantity that is proportional to the comoving number density of LLSs 
times their physical cross section \citep[e.g.][]{bah69}. In this transformation, 
\begin{equation}
dX = \frac{(1+z)^2 dz}{\sqrt{\Omega_\Lambda + \Omega_{\rm m}(1+z)^3}}\:,
\end{equation}   
and given that $dz/dX=0.27$ at $z=2.8$, we find $\ell(X)_{\tau \ge 2} = 0.33 \pm 0.08$ and  
$\ell(X)_{\tau \ge 2} = 0.51 \pm 0.13$ for the non-color and color selected samples, respectively.
In Figure \ref{lofxtau2}, the results from our survey (blue square and red cross) 
are compared to previous determinations of $\ell(X)$ in 
samples of $\tau \ge 2$ LLSs that have been selected within narrow intervals of redshift 
\citep{pro10,rib11,ome12}. Here, we also show the predicted $\ell(X)$ at $z=2.8$ 
(grey X) that we compute from an extrapolation of the \citet{ome12} $f(N_{\rm HI},X)$
assuming $\ell(z) \sim (1+z)^{1.5}$. The almost perfect agreement, consistent 
with what we previously found in the mean free path evolution, validate our previous 
assumption in constructing a model for the Lyman$-\alpha$ forest (see Section \ref{sec:mfp}).

Finally, we estimate the contribution of $\tau \ge 2$ LLSs  to 
the observed $\lambda^{912}_{\rm mfp}$. By definition,  
\begin{equation}\label{deflx}
\ell(X) = \frac{c}{H_0} n_{\rm c}(z) \phi(z)
\end{equation}   
with $n_{\rm c}$ the comoving number density of LLSs and $\phi$ their physical cross section. 
At $z\sim 2.8$, the typical separation between LLSs therefore becomes  \citep[e.g.][]{rib11} 
\begin{equation}
\lambda_{\rm lls} = \frac{c}{H_0(1+z)^3\ell(X)} \sim 235~\rm Mpc\:. 
\end{equation}   
Compared to $\lambda^{912}_{\rm mfp}= 100 \pm 29 ~\rm h^{-1}_{70.4}~Mpc$, it follows that 
systems with $N_{\rm HI} \ge 10^{17.5}\rm cm^{-2}$ account for only $40\%$ of the 
observed mean free path. Similarly, integrating Equation \ref{tauanalyt} between $z_{\rm qso}=3.01$
and $z_{912}=2.651$, we find that $\tau \ge 2$ LLSs contribute to $43\%$ of the 
total Lyman limit optical depth for the assumed \citet{ome12} $f(N_{\rm HI},X)$.
All together, our analysis reveals properties for the mean free path and 
for the number of $\tau \ge 2$ LLSs that are consistent with previous determinations at other redshifts, 
implying a smooth evolution of the optically thick gas distribution 
across $\sim 9$ Gyr of cosmic history.

\section{Physical properties of Lyman limit systems}\label{complls}

In this section, we exploit the quality of our MagE spectra
to investigate the physical properties of $\tau \ge 2$ LLSs. Here, we 
focus our analysis on a composite spectrum of LLSs to qualitatively explore 
typical values of metallicity and ionization parameters in a representative ensamble 
of systems that are selected purely based on their hydrogen column density between 
$2.6 \le z \le 3$. A detailed analysis of the physical conditions in individual LLSs 
is deferred to future work. 

In the previous sections, we showed how the number of LLSs found in 
color selected samples exceeds the number of intervening systems along 
non-color selected quasars. As discussed, this bias is due to the fact that absorption at 
the Lyman limit reddens the colors of these quasars, that are preferentially selected for spectroscopic 
follow-up because they separate from the stellar locus. Additional reddening may come from dust \citep{vla06,kap10}, 
but studies of damped Lyman-$\alpha$ systems 
(DLAs), at even higher column densities of $N_{\rm HI}\ge10^{20.3}\rm cm^{-2}$, suggest only modest 
differences in the reddening of optical and radio selected quasars with intervening DLAs \citep{jor06}.
Thus, given two LLSs along a color selected quasar and a non-color
selected quasar, there is no current indication that they will have largely 
different physical properties (e.g. their metal content), except perhaps an higher $N_{\rm HI}$ column density
for the systems along color selected quasars. For this reason,
in the remainder of this work, we combine the two populations of LLSs detected 
along color and non-color selected samples.

\begin{figure*}
\centering
\includegraphics[scale=0.6,angle=90]{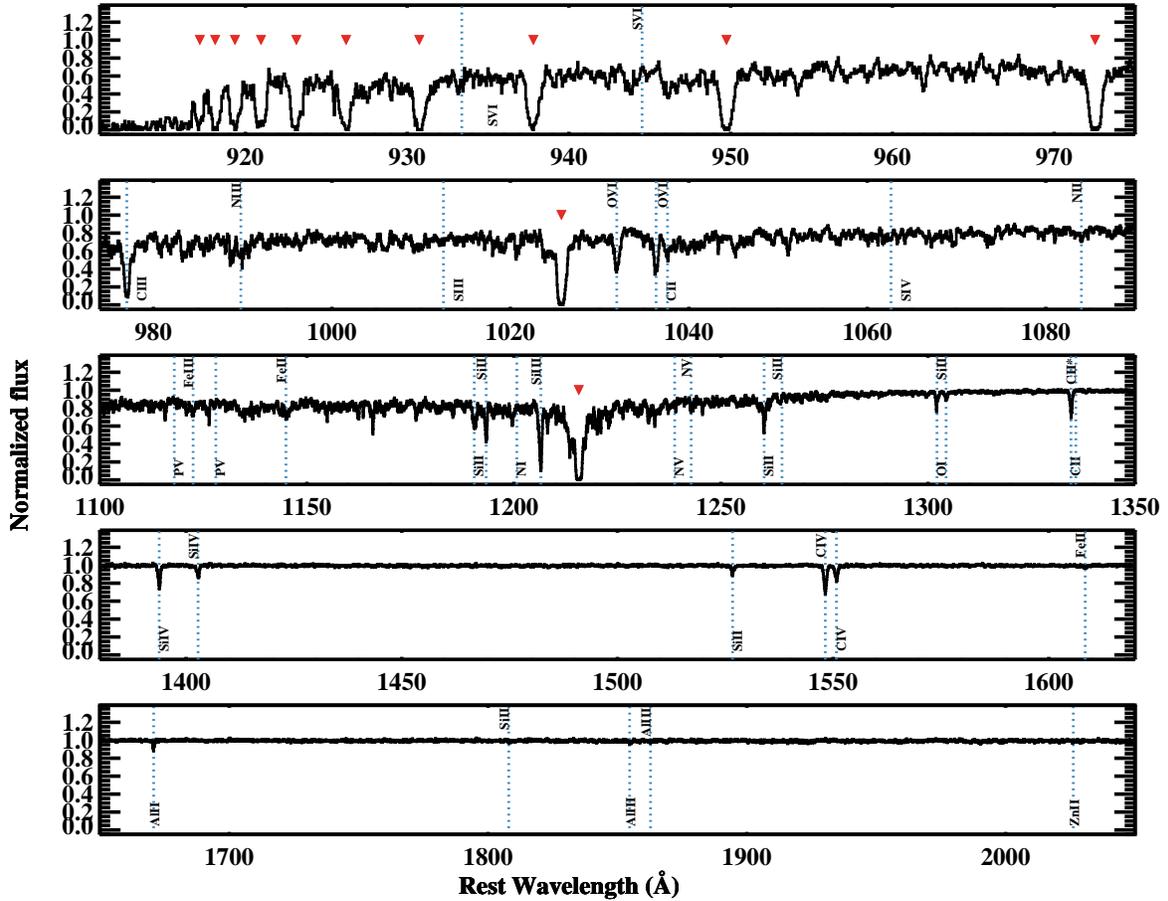}
\caption{Composite spectrum of 38 $\tau \ge 2$ LLSs which are included in our statistical sample 
  at $z \ge 2.6$. Hydrogen lines in the Lyman series are detected through Ly13 and are 
  marked by red triangles. The position of common metal absorption lines is 
  also highlighted with blue dotted vertical lines. Spectra of LLSs are characterized by 
  multiple absorption lines 
  associated to metals in different ionization states, most notably \ion{C}{3} and \ion{Si}{3}.}\label{llsstack}
\end{figure*}

\begin{figure}
\centering
\includegraphics[scale=0.34,angle=90]{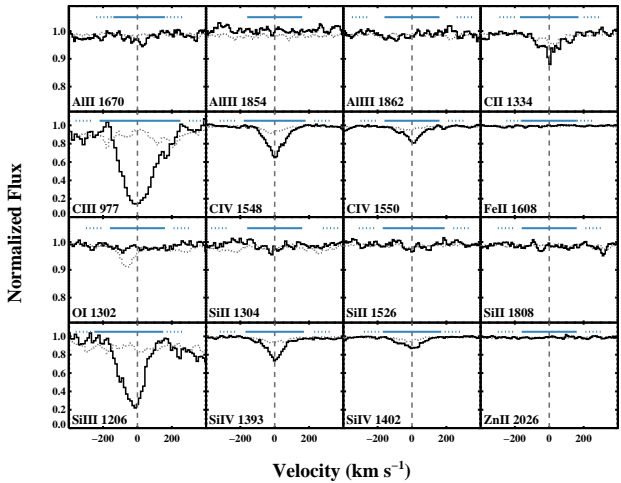}
\caption{Velocity plot of selected metal lines that are included in the photo-ionization 
  modeling of a composite spectrum of 20 LLSs with $\log N_{\rm HI}=17.5-19.0$. In grey, we show 
  the associated errors, which we compute via bootstrapping. The solid and dotted blue lines 
  at the top of each panel mark the velocity windows used to compute the line equivalent 
  widths and the local continuum, respectively. At $\log N_{\rm HI}<19.0$, only doubly and 
  triply ionized metal ions are most evident.}\label{ewplot}
\end{figure}

\subsection{The composite LLS spectrum}

To generate a composite spectrum of $z\sim 2.8$ LLSs, we first continuum normalize the 
quasar spectra by fitting a polynomial function to the data redward to the quasar 
Lyman$-\alpha$ emission lines, omitting regions where metal lines are detected in absorption. 
At wavelengths blueward to Lyman$-\alpha$, instead, we generate a continuum model using a spline 
function that is constrained by the data in regions free from evident absorption arising 
in intervening systems. Between the quasar Lyman$-\alpha$ and Lyman$-\beta$ emission lines, 
we further adopt a first guess for the continuum level using principle component analysis, 
following the procedure described in \citet{par11}.
Next, we median combine the normalized spectra in the rest-frame of the 38 $\tau \ge 2$ 
LLSs that are part of our statistical sample at $z\ge 2.6$. In this 
case, a median is preferred over the arithmetic mean to attempt to preserve 
the definition of optical depth associated to each absorption line. 
As previously done for the quasar composite spectrum, the error array is derived via bootstrapping. 

The LLS composite spectrum is shown in Figure \ref{llsstack}, 
where we mark the position of metal transitions that are common in intervening systems.
Besides the prominent hydrogen Lyman series, multiple transitions in different ionization states 
are detected, including \ion{O}{6},  \ion{Si}{4} and \ion{C}{4}, strong 
\ion{C}{3} and \ion{Si}{3}, singly ionized Si, C, Al, and \ion{O}{1}. Similar absorption lines
have been reported in a composite spectrum of LLSs at $z\sim 3.5$  \citep{pro10}.
Although systems with $N_{\rm HI}\ge 10^{19}~\rm cm^{-2}$ are formally classified as
LLSs, several studies have already investigated the typical metallicity and ionization state of 
DLAs with $N_{\rm HI} \ge 10^{20.3}~\rm cm^{-2}$ 
and of sub-DLAs or super-LLSs (SLLSs) with $N_{\rm HI} = 10^{19} - 10^{20.3}~\rm cm^{-2}$
\citep[e.g.][]{des03,raf12}. Conversely, with the exception of the pioneering study of 
\citet{ste90}, the metal properties and ionization states of $z\sim 3$ LLSs with 
$N_{\rm HI} = 10^{17.5} - 10^{19}~\rm cm^{-2}$ have been investigated in only a handful of 
systems \citep[see appendix of][]{fum11b} and a detailed study of a homogeneous 
and unbiased sample of LLSs is at present lacking. 

To explore the metal absorption properties of this sub-class of LLSs, we generate a 
second composite spectrum for the 20 $\tau \ge 2$ LLSs in our statistical samples that 
exhibit an \ion{H}{1} column density $<10^{19}~\rm cm^{-2}$, as inferred from the absence of 
damping wings in the Lyman-$\alpha$ line profile. 
In Figure \ref{ewplot}, we show the line profiles of common metal transitions in this second 
composite spectrum at $\lambda > 1300$\AA. Other transitions are detected at even shorter 
wavelength, but because of the limited sample size and the increasingly high contamination 
from the Lyman-$\alpha$ forest blueward to $\lambda = 1300$\AA, blends significantly 
affect our ability to reliably measure equivalent widths. However, 
given the predominance of doubly ionized transitions in LLSs 
\citep[see Figure \ref{llsstack} or, e.g. ][]{rib11b} and the fact that
\ion{C}{3} $\lambda 977$\AA\ and \ion{Si}{3} $\lambda 1206$\AA\  
are clearly detected in the stack, we also include these transitions 
in our analysis. The most evident feature of Figure \ref{ewplot}
is that, compared to the stack of all LLSs, the singly ionized species and \ion{O}{1} are 
mostly undetected. A similar dichotomy between lower and higher ionization lines
at different densities has been reported by \citet{pie10} in a composite spectrum of 
strong Lyman-$\alpha$ forest absorbers, with $N_{\rm HI} \gtrsim 10^{15.4}~\rm cm^{-2}$.

To characterize the strength of the selected metal lines in the composite spectrum, we measure 
the line equivalent widths $W$ within a velocity window that is large enough to encompass 
the entire line profile. For the undetected lines, we fix instead the velocity window at 
$\pm 160~\rm km~s^{-1}$, comparable to the width of the detected lines.
A summary of these measurements is provided in Table \ref{tabew}, where  
we list $3\sigma$ limits for non detections.

Given the suggested connection between LLSs and the halo gas in the surroundings of 
galaxies (see Section \ref{discussion}), we compare the typical 
equivalent widths observed in LLSs and in the circumgalactic medium (CGM) of galaxies  
at similar redshifts. Using a composite of galaxy spectra at low resolution, \citet{ste10} 
have derived the typical equivalent widths of \ion{C}{2}, \ion{Si}{4}, and \ion{C}{4} in the CGM 
of $z\sim 2-3$ galaxies. The equivalent widths for LLSs are comparable 
to the values reported for the CGM only at the largest impact parameters, 
beyond $\sim 80-100$ kpc from the galaxy centers. At face value, this result 
implies that the majority of LLSs with $N_{\rm HI} = 10^{17.5} - 10^{19}~\rm cm^{-2}$ do 
not arise from the innermost CGM of Lyman break galaxies (LBGs) with $L\gtrsim 0.3L_*$. However, 
this seems in contradiction to the empirical result that the halos of LBGs contain enough neutral 
gas to account for a significant fraction of the population of $\tau \ge 2$ LLSs 
within $\sim$100 kpc from the galaxy centers \citep{ste10}. This tension reveals a still 
incomplete understanding on the gas distribution around LBGs in connection to 
the population of LLSs, a gap that future studies need to address.

\begin{figure}
\centering
\includegraphics[scale=0.48]{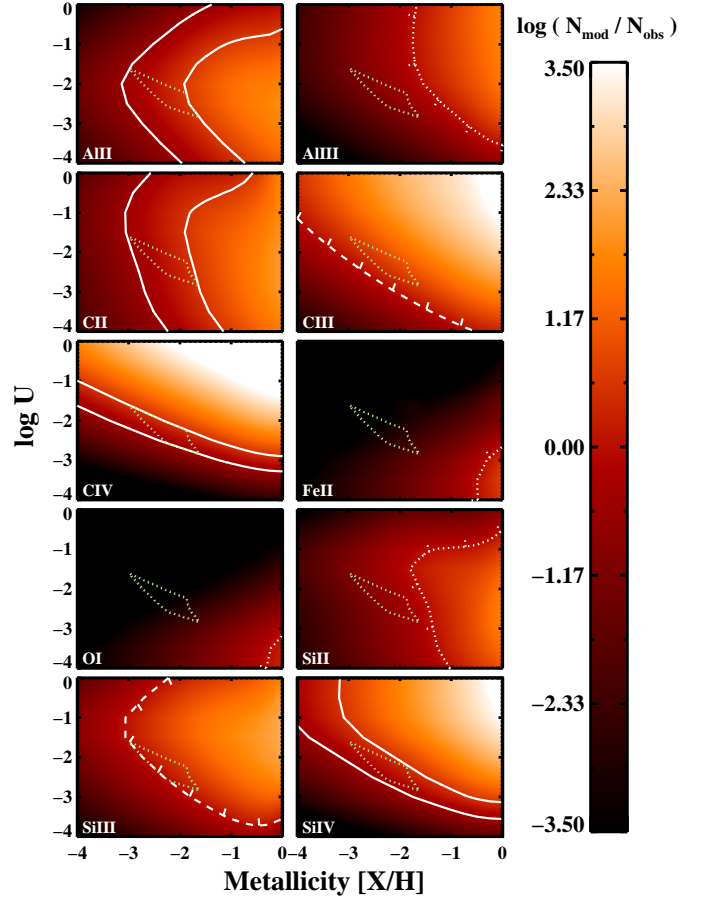}
\caption{Results from the photo-ionization modeling of the composite spectrum of 
$\log N_{\rm HI}=17.5-19.0$ LLSs. In each panel, we compare the observed line strength of
selected metal transitions to model predictions at different metallicities and ionization 
parameters. For each ion, the range of parameters that are consistent with the 
observed equivalent widths is marked by two solid white lines. For lower and upper limits, 
we show instead the allowed parameter space with dashed and dotted lines, respectively. 
The green dotted lines enclose the typical metallicity and ionization parameter that 
qualitatively reproduce the metal line strengths in the composite spectrum. 
Note that the left edge of this region is formally a lower limit.
Our analysis reveals a population of ionized and likely metal poor 
systems.}\label{cldymet}
\end{figure}

\subsection{Photo-ionization modeling}

In this section, we use photo-ionization modeling to 
qualitatively investigate the typical metallicity and ionization conditions 
that are required to reproduce the equivalent widths seen in the composite spectrum of
LLSs with $\log N_{\rm HI}=17.5-19.0$ at $z\sim2.8$. 

As evident from Figure \ref{ewplot}, the majority of the metal 
transitions in this composite spectrum lie on the linear portion of the curve of growth, with 
the exception of \ion{C}{3} and \ion{Si}{3} that are most likely saturated at the spectral 
resolution of MagE. We can therefore directly map the observed 
equivalent widths into column densities, as listed in Table \ref{tabew}. 
Consistent values of column densities are also found via the apparent 
optical depth method \citep{sav91}. Before proceeding, we emphasize that several effects 
confound the physical interpretation of column densities measured in composite spectra, 
including contamination from unrelated absorption lines or offsets in velocity between 
metal transitions and the systemic redshifts measured from hydrogen lines. 
Moreover, differential dust depletion and variation in 
the radiation field to which individual systems are exposed \citep[see e.g.][]{coo13} 
complicate the interpretation of the measured column densities.
Therefore, the listed column densities will be used only as proxy for the equivalent widths to gain 
qualitative insight on the physical properties of $z\sim 2.8$ LLSs. A detailed analysis 
of individual systems will be presented in future work.

Results from the photo-ionization modeling are summarized in Figure \ref{cldymet}, 
where we show the column densities of metal transitions as a function of $U$ and 
$\rm [Fe/H]$, normalized to the observed column densities or the corresponding limits
listed in Table \ref{tabew}. Models are computed using the {\sc cloudy} (v10.00) photo-ionization 
code \citep{fer98}, including both collisional ionization and photo-ionization at equilibrium 
and the \citet{haa12} extragalactic UV background at $z\sim 2.8$. During our analysis, 
relative abundances of different elements are fixed to the solar values \citep{asp09}.
We also assume a typical hydrogen column density of $\log N_{\rm HI} \sim 17.8$, a value that  
is consistent with the median column density derived in random realizations of 20 LLSs that 
are drawn from the \citet{ome12} $f(N_{\rm},X)$ at $z\sim 2.8$ between $\log N_{\rm HI}=17.5-19.0$. 
Further, the flux transmitted blueward to the Lyman limit is also consistent with a median 
$\log N_{\rm HI} \ge 17.8$.

Figure \ref{cldymet} reveals how the doubly and triply ionized C and Si 
transitions provide the most stringent constraints for the $U$ parameter, but they are 
mostly insensitive to the metallicity of the gas. Conversely, \ion{C}{2} and \ion{Al}{2} are 
the most constraining transitions for the metallicity, although degenerate with respect to 
$U$. Combining all the independent constraints, we conclude that 
the equivalent widths observed in the composite spectrum of $\log N_{\rm HI}=17.5-19.0$ LLSs 
can be reproduced by gas that is significantly ionized ($\log U \gtrsim -3$),
and likely metal poor ($\rm [Fe/H] \lesssim -1.5$) \citep[cf.][]{ste90,pro99}. 
We should again emphasize that the use of the median composite spectrum to describe the 
entire population of LLSs is rather simplistic, as it does not capture information 
on the shape of the distribution. This is especially relevant given the recent 
claim by \citet{leh13} of a bimodal distribution of the metallicity in LLSs at $z\lesssim 1$, 
with one population peaking around $\rm [X/H]\sim -1.5$ and a second population centered at 
$\rm [X/H]\sim -0.3$. 

Our qualitative analysis suggests that LLSs are probably more enriched than the 
IGM at $\rm [C/H]\sim -3$ \citep[e.g.][]{sim11}, and possibly enriched to $\rm [X/H]\sim -1.5$,
the median metallicity observed in DLAs \citep[e.g.][]{raf12}.
Assuming a general picture of an increasingly higher metallicity in regions 
that are progressively closer to the site of star formation, where metals are produced,
the enhanced metallicity of LLSs places this population of 
absorbers in proximity to or within galaxies halos that have been at least in part 
enriched by metals ejected from galaxies \citep[e.g.][]{she13}.

\section{Lyman Limit systems and galaxy halos}\label{discussion}

\begin{figure}
\centering
\includegraphics[scale=0.32, angle=90]{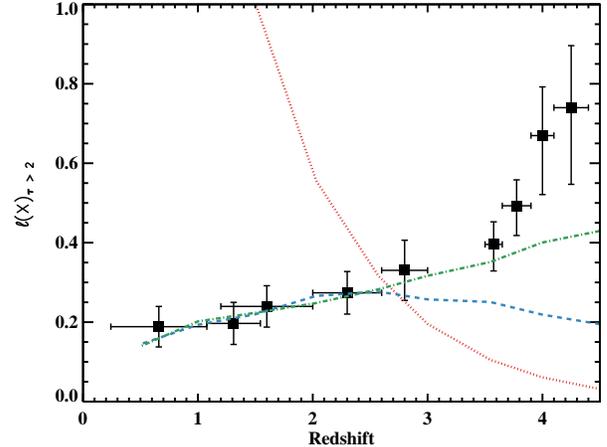}
\caption{Models for the observed evolution of $\ell(X)$ for $\tau \ge 2$ LLSs (black squares). A 
``minimal model'' in which halos within a redshift-independent mass range contribute to the number  
of LLSs proportionally to their virial area is shown with a red dotted line.
This model drastically fails to reproduce the observed $\ell(X)$. The ``wind model'' composed
by halos above a critical mass at which ejecta from supernovae do not escape the
halos is shown with a blue dashed line. This model can account for observations 
at $z\lesssim 3$ only under the ansatz that the cross section of the circumgalactic 
medium is redshift independent. 
The ``accretion model'' composed by halos that are rare fluctuations in the density field
is superimposed with a green dash dotted line. This model describes the
observed $\ell(X)$ up to $z \sim 3.5$ without any prior assumption on the 
size of the circumgalactic 
medium.}\label{modlls}
\end{figure}

In this last section, we expand the discussion of the connection between galaxies and LLSs,
to which we alluded in the previous sections. 
Already from the first quasar surveys, it became evident how LLSs differ from 
the Lyman-$\alpha$ forest \citep{tyt82,sar89,ste90} and that these absorbers are 
likely connected to galaxies. This idea has been subsequently reinforced by modern studies of the CGM 
\citep[e.g.][]{ste10,che10,pro11,rud12,wer12,chu13,pro13,sto13} that show how cold, enriched, 
and partially ionized gas from which LLSs can originate 
fills galaxy halos both in the local and distant Universe. 
At the same time, several theoretical studies have addressed the connection 
between LLSs and galaxies in the high redshift Universe 
\citep[e.g.][]{kat96,gar01,koh07,fau11,fum11,alt11,van12,fum13,rah13}. Consistent findings are that 
LLSs arise from galaxy halos, in a mixture of infalling, static, and outflowing gas.

Having a complete view of the redshift distribution of LLSs 
across nearly $\sim 10$ Gyr of cosmic history
(Figure \ref{lofxtau2}), we explore with simple toy models 
how the evolution of $\ell(X)$ relates to the evolution of the gas within 
galaxy halos. Similar investigations have been carried out in the literature by several authors.
Recently, \citet{rib11} adopted
an empirical approach to describe the evolution of LLSs in terms of the galaxy 
luminosity function and size-luminosity relation. These authors concluded 
that to account for the observed redshift dependence of $\ell(X)$, galaxies 
with $L<L_*$ have to contribute to the population of LLSs and that the physical cross section
of the absorbing material rapidly evolves from $z\sim5$ to $z\sim 2$, remaining relatively
constant afterwards. While instructive and empirical, this calculation does not provide 
a direct link between the halos in which galaxies resides and the incidence 
of LLSs. In this work, we revise this open question and propose possible scenarios 
for the co-evolution of LLSs and galaxy halos that is empirically traced
by the redshift evolution of $\ell(X)$.

We construct a simple toy model of the Universe which we populate with dark matter halos with 
comoving number density $n_{\rm gal}(M_{\rm vir},z)$. In our calculation, we assume a 
Sheth-Tormen halo mass function \citep{she99} computed as described in \citet{dek06} and corrected to 
match the results of the Bolshoi simulation \citep{kly11}. We then assign to each halo a 
characteristic virial radius $R_{\rm vir}(M_{\rm vir},z)$, using redshift dependent scaling relations 
derived from the {\sc rockstar} halo catalogues
\citep{beh13a,beh13b} extracted from the Bolshoi simulation. 
Following the definition of $\ell(X)$ in Equation (\ref{deflx}), we compute the 
galaxy contribution to the observed  $\ell(X)$ as
\begin{eqnarray}\label{modelx}
\ell(X) = \frac{4c\pi}{H_0}\int^{\log M_{\rm up}}_{\log M_{\rm low}} R_{\rm vir}^2(M_{\rm vir},z) 
f_{\rm c}(M_{\rm vir},z) \nonumber \\
 \frac{dn_{\rm gal}}{d\log M_{\rm vir}}(M_{\rm vir},z) d\log M_{\rm vir}
\end{eqnarray}
with $\log M_{\rm up}$ and $\log M_{\rm low}$ the upper and lower mass limits and 
$f_{\rm c}$ the covering factor of $\tau\ge2$ LLSs within twice the virial radius. 
Despite this formal definition, for a given cosmology, $\ell(X)$ is only constraining 
$f_{\rm c}\times R_{\rm vir}^2$, and it should be noted that 
we cannot discriminate between, e.g., a 10\% covering factor within twice the 
virial radius and a 40\% covering factor within one virial radius.  

In this model, while $M_{\rm vir}$ and $R_{\rm vir}$ are fixed by the adopted cosmology, 
$\log M_{\rm up}$, $\log M_{\rm low}$, and $f_{\rm c}$ are free parameters that 
may vary as a function of redshift and halo mass. The intrinsically one-dimensional 
nature of absorption line surveys is currently unable to disentangle the relative 
contribution of these parameters and, 
for this reason, we restrict to a minimal set of variables. First, we fix 
the upper limit $\log M_{\rm up} = 13$ to mimic the formation of 
a hot halo above a critical halo mass at which virial shocks become stable 
\citep[e.g.][]{dek06}. This limit is higher than the typical value quoted for this 
critical halo mass (between $\sim 10^{11-12}~\rm M_\odot$) to account for recent observational 
results which show a substantial amount of cold gas at or even beyond the mass scale 
at which hot halos are postulated to develop \citep{pro13,chu13}. However, because of 
the low number density of high mass halos, our result is largely insensitive to 
the assumed upper mass limit above $\log M_{\rm up} \sim 12.5$. The second simplifying 
assumption is that, lacking strong priors for the functional form of $f_{\rm c}(M_{\rm vir},z)$, 
we assume a constant covering factor both in redshift and within the mass range specified 
by $\log M_{\rm up}$ and $\log M_{\rm low}$. Thus, in our toy model, $f_{\rm c}$ merely sets 
the absolute normalization of the predicted $\ell(X)$. Validation of this assumption awaits 
future theoretical and observational work.

The first model we construct, which we dub ``minimal'', assumes that all the halos in the mass range 
$\log M_{\rm up}=11$ $\log M_{\rm low}=13$ contribute to the observed $\ell(X)$ proportionally to their
virial area ($A_{\rm vir} \propto R^2_{\rm vir} \propto M^{2/3}_{\rm vir}$). 
As shown in Figure \ref{modlls}, a choice of $f_c \sim 0.15$ at twice the virial radius 
ensures that this minimal model (red dotted line) matches the observed incidence of LLSs around 
$z\sim 2.5$. This covering factor lies in between theoretical predictions at 
comparable redshifts \citep{fum11,fau11,she13,fum13} and what is inferred around 
LBGs \citep{rud12} and quasar host galaxies \citep{pro13}.

The most evident problem with the minimal model is a complete inconsistency in the redshift 
evolution of $\ell(X)$. This failure cannot be resolved with simple scaling of $f_{\rm c}$ and originates from 
an increase of almost a factor of $\sim 20$ in the virial area at fixed halo mass between 
$z\sim 0.5-5.5$, combined with an increase of a factor of $\sim 10$ in the number density of 
halos across a comparable redshift range. In principle, this tension could be resolved by assuming 
that LLSs do not co-evolve with the properties of halo gas. However, given the 
discussion presented in the previous sections, we believe that this explanation is 
unlikely, and attribute instead the mismatch between observations and the minimal model
to having assumed that LLSs trace the CGM of a galaxy population
with fixed halo mass at all redshifts. In the following, we present two possible ways to 
resolve the failure of the minimal model, which require either a rapid evolution in the 
typical mass interval from which LLSs arise, or that the size of the optically thick region 
around galaxies is decoupled from the size of the virial region.

In the second model, we attempt to overcome the shortcomings of the minimal model
by introducing a redshift dependent lower mass limit. If we postulate that the majority of LLSs 
are associated to gas that is deposited in the halo by a feedback mechanism, we can naturally introduce 
a lower limit $\log M_{\rm low}$ by assuming that halos with circular velocities below 
$\sim 100~\rm km~s^{-1}$ \citep{dek86} are not able to retain a significant fraction 
of the ejected gas. For this reason, we define this second toy model as ``wind'' model.
This additional constraint naturally leads to a mild evolution in the 
number density of halos that host LLSs between $z\sim 0.5-5$, accounting for the factor of 
$\sim 10$ excess encountered in the minimal model. More specifically, the lower mass limit 
evolves from $\log M_{\rm low} \sim 10.5$ at $z\sim6$ to $\log M_{\rm low}\sim 11.5$ at 
$z\sim 0.5$. Due to the increasing amplitude of the halo mass function as a function of time, 
the integrated number density of halos that contribute to LLSs slightly increases at first 
with redshift. By $z\sim 2.5$, instead, due to the progressively higher value 
of  $\log M_{\rm low}$, the number density slightly declines with time.

However, the rapid increase in the virial area as a function of redshift would still cause this model 
to significantly overproduce LLSs at low redshifts, violating the observed $\ell(X)$. 
A better match to the observations is found if, instead, we fix the size 
of the CGM to the virial area at $z \sim 2.5$. Thanks to this ansatz, and by imposing a covering 
factor of  $f_c \sim 0.15$, the wind model can account for the observed $\ell(X)$ evolution 
between $z\sim 0.5-3$ (blue dashed line in Figure \ref{modlls}). 
Having imposed a constant cross section for the CGM over the last $\sim 10$ Gyr of cosmic evolution, 
this model has only limited predictive power. Nevertheless, it is interesting to note that 
similar conclusions of an invariant size of the CGM have been reached independently  
by other studies of absorption line systems \citep[e.g.][]{che12,pro13}.

Given the multiple parameters that define $\ell(X)$ in Equation \ref{modelx}, other models that 
reproduce the observed incidence of LLSs can be constructed with a different set of assumptions.
For instance, a model in which $\log M_{\rm low}$ rapidly increases with decreasing 
redshifts would naturally compensate the growth of the virial area by imposing that
LLSs arise from progressively more massive and rarer halos at later times. 
Conveniently, cosmology offers a way to impose a rapid 
redshift evolution for $\log M_{\rm low}$. If we postulate that, as suggested by 
simulations \citep{fum11,van12}, a significant fraction of LLSs arise from dense streams of 
infalling gas that penetrate inside halos, we can construct 
a third ``accretion'' model where $\log M_{\rm low}$ is set by a characteristic 
halo mass $M_*$ above which halos are significant fluctuations in the
density field. Below this critical mass, halos reside within individual filaments
of the cosmic web, where densities are comparable to the virial densities 
\citep{dek06}. Therefore, below this mass threshold, we postulate that the 
typical gas densities are below what is required to form LLSs. 

By imposing the condition $\log M_{\rm low} \ge \log M_*$, 
the accretion model (green dash dotted line in Figure \ref{modlls})
accounts for the observed redshift evolution of LLSs, without requiring any 
adjustment in the spatial extent of the CGM. In this model, however, the inclusion of low mass halos 
(down to $M\sim M_*$) forces us to adopt a lower covering factor than what was assumed 
in previous models ($f_{\rm c}\sim 0.1$ within virial radius). While the actual value of $f_{\rm c}$ 
is likely higher at most redshifts, this model provides a qualitative illustration on how the 
rapid redshift evolution of $M_*$ relegates LLSs to the CGM of progressively rarer halos,
thus compensating for the large increase in the virial area at fixed halo mass with cosmic time.  

Finally, Figure \ref{modlls} further illustrates that none of the proposed toy models is able 
to reproduce the steep rise in $\ell(X)$ beyond $z\sim3.5$. Clearly, a sudden increase in the extent 
of the CGM for $z\gtrsim 3.5$ would reproduce the observed trend \citep{rib11}. 
However, it is unclear why $z\sim 3.5$ should represent a special epoch in the evolution of the CGM. 
Instead, we argue that the different redshift evolution that is observed for $\ell(X)$ is
associated to an increasingly higher contribution of the IGM to the population of LLSs.
This is because at progressively higher redshifts the overdensities at which LLSs form approach
and exceed the halo virial density and, furthermore, the extragalactic UV background decreases. 

A simple calculation can be used to illustrate this argument. For clouds illuminated by UV 
background radiation, hydrogen densities around $\sim 0.01~\rm cm^{-3}$ are required 
for the gas to retain $>1\%$ of neutral hydrogen \citep[e.g. Appendix B of][]{fum11}. 
This threshold is comparable to the virial densities of halos at $z \sim 2.5$, while by redshift $z \sim 4$, 
typical virial densities equal $\sim 0.03~\rm cm^{-3}$, the value at which the neutral fraction approaches 
$30\%$ for a constant UV background. A similar picture is also supported by detailed radiative 
transfer calculations in numerical simulations that are able to reproduce the abundance of 
LLSs between $z\sim 3-6$  \citep{mcq11}.

\section{Summary and Conclusions}\label{summary}

In this paper, we have presented results of a survey of 105 quasars at $z\sim 3.0 \pm 0.2$ 
conducted with the blue sensitive MagE spectrograph at the Magellan Clay  telescope. 
The high signal-to-noise and 
moderate resolution data have been analyzed to investigate the properties of optically-thick hydrogen 
at $z \sim 2.6-3$, a redshift interval that has been not studied in a systematic way 
by previous surveys. Our principal findings can be summarized as it follows. 

\begin{itemize}

\item[--] By modeling the composite spectrum of quasars that are selected independently of 
their optical colors, we have estimated the mean free path of ionizing photons to be
$\lambda^{912}_{\rm mfp} = 100 \pm 29 ~\rm h_{70.4}^{-1}~ Mpc$ in the $z \sim 3$ Universe. 
This value, together with previous determinations at different redshifts 
from the literature, favors a redshift evolution of in the amplitude of the column density 
distribution function in the form $\sim (1+z)^{1.5-2}$, and implies that the Universe becomes 
transparent to ionizing radiation around redshift $z\sim 1.4-1.8$. During this analysis, we have also 
confirmed the presence of a previously discovered selection bias in SDSS that systematically 
misses blue quasars in proximity to the stellar locus. 

\item[--] We have conducted a survey of $\tau \ge 2$ LLSs in the wavelength range 
$\lambda\sim 3200-3800$ \AA\ and we have characterized the number of LLSs per unit redshift  
$\ell(z)= 1.21 \pm 0.28$ at $z\sim 2.8$. For the adopted cosmological model, this quantity 
translates into $\ell(X)= 0.33 \pm 0.08$. Our new determination is fully consistent with previous 
measurements at lower and higher redshifts, implying a smooth evolution of $\ell(X)$ between 
$z\sim 4.5$ and $z\sim 0.5$.  Combined with our measurement of the mean free path at the 
same redshift, we have concluded that LLSs with $\log N_{\rm HI} \ge 17.5~\rm cm^{-2}$ contribute 
to $\sim 40\%$ of total mean free path at $z\sim 3$.

\item[--] With the aid of photo-ionization modeling, we have inferred that a population of 
ionized ($U\gtrsim -3$)  and likely metal poor ($\rm [X/H] \lesssim -1.5$) LLSs is required
to reproduced the metal line equivalent widths observed in a composite spectrum of 20
LLSs with $\log N_{\rm HI} \sim 17.5 - 19~\rm cm^{-2}$ at $z\sim2.8$. 

\item[--] We have proposed three simple toy models to describe the redshift evolution of $\tau \ge 2$
LLSs in connection to galaxies. Our main finding is that a model in which LLSs arise in halos of fixed
mass at all redshifts drastically fails to reproduce the observed number of LLSs per unit redshift. 
Successful models require either a rapid redshift evolution of the minimum mass at which LLSs form or 
a redshift invariant spatial extent of the CGM in the last $\sim 10$ Gyr of cosmic evolution.
In all cases, however, the rapid evolution observed in $\ell(X)$ at $z\ge 3.5$ is not accounted 
for, and we speculate that an increasingly higher contribution from gas within the IGM is required. 
\end{itemize}

In conclusion, our MagE survey of $z\sim 3$ quasars has provided the missing pieces in 
a consistent picture of the cosmological evolution of optically thick gas
between $z\sim 0.5-4.5$. The redshift evolution of the mean free path of ionizing radiation and the
incidence of $\tau \ge 2$ LLSs are now well characterized with homogeneous
techniques across $\sim 10$ Gyr of cosmic history, albeit with some controversy at $z\sim 2.5$. 
Our survey has also provided a homogeneous sample of hydrogen-selected LLSs that can be 
exploited to characterize the physical properties these absorbers and their connections to 
galaxies at $z\sim 2.6-3$. 
The analysis presented in this paper has offered only a first glimpse of the typical 
metallicity of LLSs, together with viable ways to model the observed evolution 
of $\ell(X)$. Clearly, our work does not provide the final answer to some of the riddles we 
encountered along the way, including the lower equivalent widths of metal lines in 
a composite of LLSs compared to what measured in a composite spectrum of galaxies. 
The task of formulating a coherent picture for the co-evolution of galaxies and LLSs is therefore 
a challenge that future studies will have to undertake.

\acknowledgements
We wish to dedicate this work to the memory of Wallace Sargent for his great contribution  
to our modern view of LLSs in the distant universe. It is a pleasure to thank A. Dekel and 
M. Rauch for insightful comments, and A. Cucchiara for collecting 
the KAST spectra discussed in this paper. We also thank the anonymous referee for comments 
and criticisms that helped improving this paper. Support for M.F. was provided by NASA through 
Hubble Fellowship grant HF-51305.01-A awarded by the Space Telescope Science Institute, 
which is operated by the Association of Universities for Research in Astronomy, Inc., 
for NASA, under contract NAS 5-26555. Some of observations presented in this paper were 
collected as part of two NOAO proposals  (2010B-0424 and 2009B-0394; PI: JMO). 
JMO also acknowledges travel support from the VPAA office 
at Saint Michael's College. JXP acknowledges support from NSF grants AST 10-10004 and
AST 11-09447. We acknowledge the use of the Sloan Digital Sky Survey (http://www.sdss.org/).

\LongTables 
\input{table1.tex}
\input{table2.tex}

\input{table3.tex}

\end{document}

%% file: table1.tex
\begin{deluxetable*}{lrrccclcccc}
\tablewidth{0pc}
\tablecaption{Logbook of the observations conducted with the MagE spectrograph.\label{qsoobs}}
\tablehead{
\colhead{Name}
& \colhead{R.A.} 
& \colhead{Dec.} 
& \colhead{$z_{\rm qso}$} 
& \colhead{$\sigma_{z,\rm qso}$\tablenotemark{a}} 
& \colhead{$\sigma_{v,\rm qso}$\tablenotemark{b}} 
& \colhead{Notes\tablenotemark{c}}
& \colhead{Date} 
& \colhead{Exposure}
& \colhead{Slit}
& \colhead{S/N at 5500\AA}\\
& \colhead{(J2000)} 
& \colhead{(J2000)} 
&  
& 
& \colhead{($\rm km~s^{-1}$)} 
& 
& \colhead{(yymmdd)} 
& \colhead{(sec)}
& \colhead{('')}
& }
\startdata 
FBQSJ0002+0021     & 00:02:21.11  &   +00:21:49.3  &   3.072  & 0.007 & 521 & NC    & 120713				 & 3000 & 0.7 &  9.7  \\
Q0038-4041         & 00:40:49.50  &   -40:25:14.0  &   2.977  & 0.009 & 714 & NC    & 120715				 & 2700 & 0.7 & 26.0  \\
LBQS0041-2638      & 00:43:42.80  &   -26:22:11.0  &   3.058  & 0.007 & 521 & NC    & 091223				 & 2000 & 0.7 & 16.0  \\
LBQS0047-3050      & 00:50:20.10  &   -30:34:21.0  &   2.967  & 0.007 & 521 & NC    & 091223				 & 1000 & 0.7 & 13.3  \\
UM669              & 01:05:16.80  &   -18:46:42.0  &   3.041  & 0.007 & 521 & NC    & 091223				 & 1000 & 0.7 & 20.1  \\
Q0130-403          & 01:33:01.90  &   -40:06:28.0  &   3.031  & 0.007 & 521 & NC    & 091222				 & 1200 & 0.7 & 30.3  \\
UM148              & 01:56:36.00  &   +04:45:28.0  &   3.001  & 0.007 & 521 & NC    & 091223				 & 1000 & 0.7 & 14.1  \\
CTS0418            & 02:17:41.80  &   -37:01:00.0  &   2.916  & 0.007 & 521 & NC    & 091222				 &  900 & 0.7 & 26.2  \\
H0216+0803         & 02:18:57.30  &   +08:17:28.0  &   3.001  & 0.010 & 714 & NC    & 100814				 & 1200 & 0.7 & 20.4  \\
CTS0220            & 02:41:22.70  &   -36:33:19.0  &   3.109  & 0.007 & 521 & NC    & 091223				 &  900 & 0.7 & 16.2  \\
Q0244-302          & 02:46:34.20  &   -30:04:55.0  &   3.088  & 0.007 & 521 & NC    & 091222				 & 1000 & 0.7 & 11.2  \\
CTS0424            & 03:16:43.80  &   -56:51:45.0  &   3.036  & 0.007 & 521 & NC    & 091222/23 			 & 2200 & 0.7 & 32.4  \\
CTS0238            & 03:30:10.30  &   -46:26:15.0  &   3.112  & 0.007 & 521 & NC    & 091222				 & 1100 & 0.7 & 15.1  \\
Q0351-3740         & 03:53:08.50  &   -37:40:54.0  &   2.952  & 0.007 & 521 & NC    & 091222				 & 1200 & 0.7 & 22.3  \\
Q0351-390          & 03:53:19.20  &   -38:55:56.0  &   3.007  & 0.010 & 714 & NC    & 091223				 &  600 & 0.7 & 23.9  \\
CTS0246            & 04:04:01.90  &   -33:35:00.0  &   3.042  & 0.007 & 521 & NC    & 091223				 &  800 & 0.7 & 20.4  \\
CTQ0247            & 04:07:18.00  &   -44:10:14.0  &   3.021  & 0.010 & 714 & NC    & 091223				 & 1000 & 0.7 & 25.6  \\
CTS0648            & 04:39:06.90  &   -50:47:40.0  &   2.947  & 0.007 & 521 & NC    & 091222				 & 1200 & 0.7 & 21.5  \\
CTS0260            & 04:45:52.00  &   -31:58:43.0  &   2.722  & 0.006 & 521 & NC    & 091222				 &  900 & 0.7 & 22.0  \\
H0449-1325         & 04:51:42.60  &   -13:20:33.0  &   3.107  & 0.007 & 521 & NC    & 091222				 & 1100 & 0.7 & 23.6  \\
CTS0269            & 05:17:42.20  &   -37:54:46.0  &   3.045  & 0.007 & 521 & NC    & 091222				 &  800 & 0.7 & 21.7  \\
Q0642-506          & 06:43:27.00  &   -50:41:13.0  &   3.107  & 0.007 & 521 & NC    & 091222				 &  900 & 0.7 & 19.9  \\
GB6J0833+1123      & 08:33:14.36  &   +11:23:36.3  &   2.986  & 0.007 & 521 & NC    & 091222/23,110329		         & 3600 & 0.7 & 33.2  \\
SDSSJ0904+1309     & 09:04:23.37  &   +13:09:20.7  &   2.975  & 0.007 & 521 & NC    & 091222/23 			 & 1800 & 0.7 & 38.1  \\
HE0940-1050        & 09:42:53.40  &   -11:04:25.0  &   3.076  & 0.007 & 521 & NC    & 091222				 &  800 & 0.7 & 36.5  \\
CTS0281            & 09:52:33.30  &   -23:53:48.0  &   2.904  & 0.008 & 631 & NC    & 091222,110329			 & 2400 & 0.7 & 28.8  \\
PKS1010-427        & 10:12:37.80  &   -42:58:37.8  &   2.962  & 0.007 & 521 & NC    & 091223				 & 1000 & 0.7 & 16.6  \\
TXS1033+137        & 10:36:26.88  &   +13:26:51.7  &   3.098  & 0.007 & 521 & NC    & 110328				 & 1200 & 0.7 & 19.9  \\
CTS0296            & 10:58:09.20  &   -26:05:39.0  &   2.887  & 0.007 & 521 & NC    & 091222				 &  900 & 0.7 & 29.8  \\
HS1200+1539        & 12:03:31.29  &   +15:22:54.7  &   2.985  & 0.007 & 521 & NC    & 110328,120711			 & 2100 & 0.7 & 41.9  \\
LBQS1209+1524      & 12:12:32.04  &   +15:07:25.6  &   3.067  & 0.007 & 521 & NC    & 110329				 & 2400 & 0.7 & 21.1  \\
Q1210-1049         & 12:12:52.10  &   -11:06:10.0  &   2.938  & 0.007 & 521 & NC    & 120712				 & 2700 & 0.7 & 17.0  \\
LBQS1223+1753      & 12:26:07.19  &   +17:36:49.8  &   2.953  & 0.009 & 654 & NC    & 110329				 & 1200 & 0.7 & 23.4  \\
SDSSJ1241+1230     & 12:41:58.18  &   +12:30:59.3  &   2.979  & 0.009 & 714 & NC    & 120711				 & 3600 & 0.7 & 12.3  \\
CTS0320            & 13:17:44.10  &   -31:47:14.0  &   2.965  & 0.007 & 521 & NC    & 110328				 & 2200 & 0.7 & 19.7  \\
LBQS1345-0120      & 13:48:16.65  &   -01:35:09.9  &   2.957  & 0.007 & 521 & NC    & 120713				 & 2700 & 0.7 & 26.0  \\
CTS0331            & 14:01:38.00  &   -13:46:10.0  &   3.022  & 0.011 & 792 & NC    & 120715				 & 2700 & 0.7 & 20.4  \\
Q1406+123          & 14:08:38.92  &   +12:07:09.6  &   2.944  & 0.007 & 521 & NC    & 120714				 & 3000 & 0.7 & 33.2  \\
Q1455+123          & 14:58:07.50  &   +12:09:37.8  &   3.057  & 0.007 & 521 & NC    & 120712				 & 3000 & 0.7 & 18.6  \\
Q1508+087          & 15:10:47.37  &   +08:35:35.2  &   2.999  & 0.011 & 792 & NC    & 120713				 & 3300 & 0.7 & 13.1  \\
Q1510+105          & 15:13:04.41  &   +10:23:30.8  &   3.064  & 0.007 & 521 & NC    & 120715				 & 3600 & 0.7 & 23.0  \\
SDSSJ1559+1631     & 15:59:36.89  &   +16:31:01.5  &   3.072  & 0.009 & 631 & NC    & 120715				 & 3600 & 0.7 & 18.0  \\
SDSSJ1604+1645     & 16:04:41.47  &   +16:45:38.3  &   2.888  & 0.010 & 792 & NC-B  & 100813				 & 2900 & 0.7 & 56.7  \\
Q1623+155          & 16:25:47.21  &   +15:27:21.1  &   3.060  & 0.009 & 631 & NC    & 120712				 & 2700 & 0.7 & 14.3  \\
PMNJ1837-5848      & 18:37:53.80  &   -58:48:09.0  &   3.041  & 0.007 & 521 & NC    & 120712/15 			 & 5700 & 0.7 &  6.2  \\
Q2117-4600         & 21:21:11.50  &   -45:47:58.0  &   2.957  & 0.007 & 521 & NC    & 120715				 & 3600 & 0.7 &  8.9  \\
Q2119-4307A        & 21:22:28.70  &   -42:54:38.0  &   2.952  & 0.007 & 521 & NC-B  & 120713				 & 3600 & 0.7 & 16.4  \\
Q2135-1335         & 21:27:45.40  &   -13:22:13.0  &   2.928  & 0.007 & 521 & NC    & 100813/14 			 & 2700 & 0.7 & 21.5  \\
FBQS2129+0037      & 21:29:16.60  &   +00:37:56.6  &   2.961  & 0.007 & 521 & NC    & 100814				 & 1100 & 0.7 & 26.3  \\
CTS0358            & 21:41:44.40  &   -38:40:41.0  &   3.102  & 0.009 & 631 & NC    & 120715				 & 3300 & 0.7 & 22.0  \\
PSGS2144+0542      & 21:47:22.26  &   +05:56:19.5  &   2.909  & 0.007 & 521 & NC    & 120715				 & 3600 & 0.7 & 12.2  \\
SDSSJ2201+1256     & 22:01:16.75  &   +12:56:36.4  &   2.927  & 0.009 & 714 & NC    & 120713				 & 3000 & 0.7 & 17.5  \\
FBQSJ2207+0101     & 22:07:08.32  &   +01:01:25.2  &   2.904  & 0.007 & 521 & NC    & 120715				 & 3000 & 0.7 & 26.2  \\
PC2211+0119        & 22:14:27.80  &   +01:34:57.0  &   3.109  & 0.007 & 521 & NC    & 120713				 & 3000 & 0.7 & 20.6  \\
LBQS2231-0015      & 22:34:08.99  &   +00:00:01.6  &   3.025  & 0.010 & 714 & NC    & 100813				 & 1200 & 0.7 & 31.3  \\
HE2243-6031        & 22:47:09.10  &   -60:15:45.0  &   3.005  & 0.011 & 792 & NC    & 100813,120712	 		 & 1500 & 0.7 & 56.1  \\
UM659              & 23:14:07.20  &   -03:25:28.0  &   3.056  & 0.009 & 631 & NC    & 120712				 & 2700 & 0.7 & 14.6  \\
PKS2314-340        & 23:16:43.39  &   -33:49:12.5  &   2.963  & 0.007 & 521 & NC-B  & 100813				 & 1200 & 0.7 & 24.2  \\
FBQS2330-0120      & 23:30:54.40  &   -01:20:53.0  &   2.910  & 0.007 & 521 & NC    & 100814				 & 1500 & 0.7 & 14.0  \\
FBQSJ2339+0030     & 23:39:30.00  &   +00:30:17.3  &   3.053  & 0.007 & 521 & NC    & 120712				 & 2700 & 0.7 & 20.1  \\
UM184              & 23:50:57.87  &   -00:52:09.9  &   3.021  & 0.010 & 714 & NC    & 120712				 & 3000 & 0.7 & 23.8  \\
SDSSJ0010-0037     & 00:10:22.17  &   -00:37:01.3  &   3.153  & 0.007 & 521 & C     & 100814				 & 1200 & 0.7 & 12.3  \\
SDSSJ0043-0015     & 00:43:23.43  &   -00:15:52.6  &   2.829  & 0.008 & 631 & C     & 100814				 & 1200 & 0.7 & 18.6  \\
SDSSJ0125-1027     & 01:25:30.85  &   -10:27:39.8  &   3.356  & 0.008 & 521 & C     & 100813/14 			 & 2400 & 0.7 & 27.8  \\
SDSSJ0139-0824     & 01:39:01.40  &   -08:24:43.9  &   3.016  & 0.010 & 714 & C     & 091223,100813/14		         & 3600 & 0.7 & 21.0  \\
SDSSJ0145-0945     & 01:45:16.59  &   -09:45:17.3  &   2.732  & 0.006 & 521 & C     & 100813				 &  900 & 0.7 & 37.0  \\
SDSSJ0145-0945B    & 01:45:16.76  &   -09:45:17.9  &   2.731  & 0.006 & 521 & C     & 100813				 & 1500 & 0.7 & 14.9  \\
SDSSJ0148-0907     & 01:48:50.64  &   -09:07:12.8  &   3.322  & 0.008 & 521 & C     & 100814				 & 1200 & 0.7 & 22.4  \\
SDSSJ0209-0005     & 02:09:50.70  &   -00:05:06.4  &   2.858  & 0.007 & 521 & C     & 100814				 &  900 & 0.7 & 37.2  \\
SDSSJ0251-0737     & 02:51:51.20  &   -07:37:07.7  &   3.084  & 0.007 & 521 & C     & 091223				 & 1000 & 0.7 & 12.9  \\
SDSSJ0303-0023     & 03:03:41.04  &   -00:23:21.8  &   3.229  & 0.007 & 521 & C     & 091223				 &  900 & 0.7 & 24.4  \\
SDSSJ0304-0008     & 03:04:49.86  &   -00:08:13.4  &   3.297  & 0.009 & 631 & C     & 091223,100814			 & 2000 & 0.7 & 34.0  \\
SDSSJ0338-0005     & 03:38:54.77  &   -00:05:20.9  &   3.053  & 0.010 & 714 & C     & 091223				 & 1000 & 0.7 & 15.7  \\
SDSSJ0859+0205     & 08:59:59.15  &   +02:05:19.7  &   2.979  & 0.009 & 654 & C     & 091222/23 			 & 2400 & 0.7 & 27.9  \\
SDSSJ0908+0658     & 09:08:32.28  &   +06:58:53.8  &   3.062  & 0.007 & 521 & C     & 110329				 & 1200 & 0.7 & 14.5  \\
SDSSJ0915+0549     & 09:15:46.68  &   +05:49:42.7  &   2.976  & 0.007 & 521 & C     & 110329				 & 1200 & 0.7 & 23.7  \\
SDSSJ0931-0000     & 09:31:53.12  &   -00:00:51.0  &   3.209  & 0.007 & 521 & C     & 091223				 & 1100 & 0.7 & 16.7  \\
SDSSJ0937+0417     & 09:37:15.38  &   +04:17:37.6  &   3.127  & 0.007 & 521 & C     & 091223				 & 1100 & 0.7 & 16.6  \\
SDSSJ0942+0422     & 09:42:02.04  &   +04:22:44.6  &   3.283  & 0.009 & 631 & C     & 091222/23 			 & 1500 & 0.7 & 40.1  \\
SDSSJ0947+1421     & 09:47:34.20  &   +14:21:17.0  &   3.041  & 0.007 & 521 & C     & 110328				 &  900 & 0.7 & 28.3  \\
SDSSJ1004+0018     & 10:04:28.43  &   +00:18:25.6  &   3.050  & 0.009 & 631 & C     & 091222/23 			 & 2100 & 0.7 & 24.9  \\
SDSSJ1015+1118     & 10:15:39.35  &   +11:18:15.9  &   2.913  & 0.007 & 521 & C     & 110329				 & 1200 & 0.7 & 18.7  \\
SDSSJ1019+0825     & 10:19:54.53  &   +08:25:15.0  &   2.989  & 0.007 & 521 & C     & 110328				 & 1200 & 0.7 & 23.7  \\
SDSSJ1025+0452     & 10:25:09.64  &   +04:52:46.7  &   3.243  & 0.007 & 521 & C     & 091223				 &  900 & 0.7 & 20.2  \\
SDSSJ1205-0048     & 12:05:32.23  &   -00:48:48.2  &   2.986  & 0.007 & 521 & C     & 110328				 & 1200 & 0.7 & 17.8  \\
SDSSJ1405+0507     & 14:05:21.67  &   +05:07:44.4  &   2.994  & 0.007 & 521 & C-B   & 110329				 & 1200 & 0.7 & 18.0  \\
SDSSJ1432+1139     & 14:32:12.86  &   +11:39:53.1  &   3.006  & 0.010 & 714 & C-B   & 110329				 & 1000 & 0.7 & 13.8  \\
SDSSJ1459+0024     & 14:59:07.18  &   +00:24:01.2  &   3.024  & 0.007 & 521 & C     & 110329				 & 1500 & 0.7 & 16.6  \\
SDSSJ1513+0855     & 15:13:52.52  &   +08:55:55.7  &   2.891  & 0.007 & 521 & C     & 100812/13 			 & 2100 & 0.7 & 29.3  \\
SDSSJ1521-0048     & 15:21:19.68  &   -00:48:18.6  &   2.949  & 0.007 & 521 & C     & 100812				 &  630 & 0.7 & 10.4  \\
SDSSJ1550+0537     & 15:50:36.80  &   +05:37:49.9  &   3.153  & 0.009 & 631 & C     & 100814				 & 1100 & 0.7 & 21.3  \\
SDSSJ1551+0908     & 15:51:03.39  &   +09:08:49.2  &   2.755  & 0.007 & 521 & C     & 100814				 & 1100 & 0.7 & 24.6  \\
SDSSJ1608+0715     & 16:08:43.90  &   +07:15:08.6  &   2.888  & 0.008 & 631 & C     & 100814				 &  900 & 0.7 & 37.9  \\
SDSSJ1615+0608     & 16:15:45.95  &   +06:08:52.4  &   3.062  & 0.007 & 521 & C-B   & 110328				 & 1200 & 0.7 & 18.6  \\
SDSSJ1620+0941     & 16:20:50.27  &   +09:41:35.2  &   3.061  & 0.007 & 521 & C     & 100814				 & 1200 & 0.7 & 14.1  \\
SDSSJ2038-0025     & 20:38:14.55  &   -00:25:38.9  &   2.707  & 0.006 & 521 & C     & 100813/14 			 & 2700 & 0.7 & 23.2  \\
SDSSJ2049-0554     & 20:49:46.35  &   -05:54:53.4  &   3.195  & 0.007 & 521 & C     & 100813				 & 1200 & 0.7 & 17.8  \\
SDSSJ2055-0511     & 20:55:16.84  &   -05:11:11.0  &   3.004  & 0.010 & 714 & C-B   & 100813/14 			 & 2400 & 0.7 & 27.6  \\
SDSSJ2100-0641     & 21:00:25.03  &   -06:41:46.0  &   3.130  & 0.007 & 521 & C     & 100814				 & 1200 & 0.7 & 17.8  \\
SDSSJ2222-0946     & 22:22:56.11  &   -09:46:36.2  &   2.914  & 0.007 & 521 & C     & 100813				 & 1200 & 0.7 & 22.8  \\
SDSSJ2225-0041     & 22:25:59.52  &   -00:41:57.5  &   2.774  & 0.007 & 521 & C     & 100814				 & 1100 & 0.7 & 22.2  \\
SDSSJ2238-0921     & 22:38:19.76  &   -09:21:06.0  &   3.278  & 0.007 & 521 & C     & 100813				 & 1200 & 0.7 & 20.9  \\
SDSSJ2319-1040     & 23:19:34.77  &   -10:40:36.9  &   3.180  & 0.007 & 521 & C     & 100813				 & 1200 & 0.7 & 17.5  \\
SDSSJ2334-0908     & 23:34:46.40  &   -09:08:12.3  &   3.351  & 0.008 & 521 & C     & 100813				 & 1200 & 0.7 & 24.0  \\
SDSSJ2348-1041     & 23:48:56.48  &   -10:41:31.2  &   3.153  & 0.007 & 521 & C     & 100813				 & 1200 & 0.7 & 20.9 
\enddata  
\tablenotetext{a}{Statistical error on the quasar redshift from broad emission lines.} 
\tablenotetext{b}{Corresponding error on the quasar redshift in velocity.} 
\tablenotetext{c}{C: color selected; NC: non color selected; B: broad absorption line quasar (visually classified)} 
\end{deluxetable*}


%% file: table2.tex
\begin{deluxetable*}{lccccc}
\tablewidth{0pc}
\tablecaption{Results of the $\tau \ge 2$ LLS survey.\label{sellls}}
\tablehead{
\colhead{Name}            &  
\colhead{Notes\tablenotemark{a}}           &
\colhead{$z_{\rm qso}$}   &
\colhead{$z_{\rm start}$} &
\colhead{$z_{\rm end}$}   &
\colhead{$z_{\rm lls}$}}  
\startdata
FBQSJ0002+0021  & NC           &  3.072 & 3.0315 & 2.5000 & 0.0000  \\  
Q0038-4041      & NC           &  2.977 & 2.9374 & 2.8161 & 2.8161  \\  
LBQS0041-2638   & NC           &  3.058 & 3.0176 & 2.5000 & 0.0000  \\  
LBQS0047-3050   & NC           &  2.967 & 2.9275 & 2.5000 & 0.0000  \\  
UM669           & NC-(1)       &  3.041 & 3.0008 & 2.9271 & 2.9271  \\  
Q0130-403       & NC           &  3.031 & 2.9909 & 2.5620 & 2.5620  \\  
UM148           & NC-(1)       &  3.001 & 2.9612 & 2.5000 & 0.0000  \\  
CTS0418         & NC           &  2.916 & 2.8770 & 2.5828 & 2.5828  \\  
H0216+0803      & NC-(1)       &  3.001 & 2.9612 & 2.5000 & 0.0000  \\  
CTS0220         & NC           &  3.109 & 3.0681 & 2.7391 & 2.7391  \\  
Q0244-302       & NC           &  3.088 & 3.0473 & 3.0473 & 3.0909  \\  
CTS0424         & NC           &  3.036 & 2.9958 & 2.5738 & 2.5738  \\  
CTS0238         & NC           &  3.112 & 3.0711 & 2.5000 & 0.0000  \\  
Q0351-3740      & NC           &  2.952 & 2.9127 & 2.5000 & 0.0000  \\  
Q0351-390       & NC-(2)       &  3.007 & 2.9671 & 2.5040 & 2.5040  \\  
CTS0246         & NC           &  3.042 & 3.0018 & 2.5000 & 0.0000  \\  
CTQ0247         & NC           &  3.021 & 2.9810 & 2.6219 & 2.6219  \\  
CTS0648         & NC           &  2.947 & 2.9077 & 2.7957 & 2.7957  \\  
CTS0260         & NC           &  2.722 & 2.6849 & 2.5000 & 0.0000  \\  
H0449-1325      & NC           &  3.107 & 3.0661 & 2.9978 & 2.9978  \\  
CTS0269         & NC           &  3.045 & 3.0047 & 2.5000 & 0.0000  \\  
Q0642-506       & NC           &  3.107 & 3.0661 & 2.6591 & 2.6591  \\  
GB6J0833+1123   & NC-(1)       &  2.986 & 2.9463 & 2.5000 & 0.0000  \\  
SDSSJ0904+1309  & NC           &  2.975 & 2.9354 & 2.7690 & 2.7690  \\  
HE0940-1050     & NC           &  3.076 & 3.0354 & 2.9169 & 2.9169  \\  
CTS0281         & NC           &  2.904 & 2.8651 & 2.4968 & 2.4968  \\  
PKS1010-427     & NC           &  2.962 & 2.9226 & 2.5000 & 0.0000  \\  
TXS1033+137     & NC           &  3.098 & 3.0572 & 2.5977 & 2.5977  \\  
CTS0296         & NC           &  2.887 & 2.8483 & 2.5432 & 2.5432  \\  
HS1200+1539     & NC           &  2.985 & 2.9453 & 2.7077 & 2.7077  \\  
LBQS1209+1524   & NC           &  3.067 & 3.0265 & 2.5000 & 0.0000  \\  
Q1210-1049      & NC           &  2.938 & 2.8988 & 2.5000 & 0.0000  \\  
LBQS1223+1753   & NC           &  2.953 & 2.9136 & 2.9136 & 2.9584  \\  
SDSSJ1241+1230  & NC           &  2.979 & 2.9394 & 2.9394 & 2.9870  \\  
CTS0320         & NC           &  2.965 & 2.9255 & 2.5000 & 0.0000  \\  
LBQS1345-0120   & NC           &  2.957 & 2.9176 & 2.8827 & 2.8827  \\  
CTS0331         & NC           &  3.022 & 2.9820 & 2.5000 & 0.0000  \\  
Q1406+123       & NC           &  2.944 & 2.9047 & 2.5000 & 0.0000  \\  
Q1455+123       & NC           &  3.057 & 3.0166 & 2.6479 & 2.6479  \\  
Q1508+087       & NC           &  2.999 & 2.9592 & 2.7222 & 2.7222  \\  
Q1510+105       & NC           &  3.064 & 3.0235 & 2.8391 & 2.8391  \\  
SDSSJ1559+1631  & NC           &  3.072 & 3.0315 & 3.0315 & 3.0611  \\  
SDSSJ1604+1645  & NC-B         &  2.888 & 2.8493 & 2.5000 & 0.0000  \\  
Q1623+155       & NC           &  3.060 & 3.0196 & 2.5000 & 0.0000  \\  
PMNJ1837-5848   & NC           &  3.041 & 3.0008 & 2.7290 & 2.7290  \\  
Q2117-4600      & NC           &  2.957 & 2.9176 & 2.5000 & 0.0000  \\  
Q2119-4307A     & NC-B         &  2.952 & 2.9127 & 2.5362 & 2.5362  \\  
Q2135-1335      & NC           &  2.928 & 2.8889 & 2.5000 & 0.0000  \\  
FBQS2129+0037   & NC           &  2.961 & 2.9216 & 2.9173 & 2.9173  \\  
CTS0358         & NC           &  3.102 & 3.0612 & 2.8932 & 2.8932  \\  
PSGS2144+0542   & NC           &  2.909 & 2.8701 & 2.5000 & 0.0000  \\  
SDSSJ2201+1256  & NC           &  2.927 & 2.8879 & 2.5000 & 0.0000  \\  
FBQSJ2207+0101  & NC           &  2.904 & 2.8651 & 2.8651 & 2.8684  \\  
PC2211+0119     & NC           &  3.109 & 3.0681 & 2.5000 & 0.0000  \\  
LBQS2231-0015   & NC           &  3.025 & 2.9849 & 2.6525 & 2.6525  \\  
HE2243-6031     & NC           &  3.005 & 2.9651 & 2.5000 & 0.0000  \\  
UM659           & NC-(1)       &  3.056 & 3.0156 & 2.5000 & 0.0000  \\  
PKS2314-340     & NC-B         &  2.963 & 2.9235 & 2.5000 & 0.0000  \\  
FBQS2330-0120   & NC           &  2.910 & 2.8711 & 2.5000 & 0.0000  \\  
FBQSJ2339+0030  & NC           &  3.053 & 3.0126 & 2.5000 & 0.0000  \\  
UM184           & NC-(1)       &  3.021 & 2.9810 & 2.9295 & 2.9295  \\  
SDSSJ0010-0037  & C            &  3.153 & 3.1117 & 3.1117 & 3.1166  \\  
SDSSJ0043-0015  & C            &  2.829 & 2.7909 & 2.5000 & 0.0000  \\  
SDSSJ0125-1027  & C            &  3.356 & 3.3126 & 2.5625 & 2.5625  \\  
SDSSJ0139-0824  & C            &  3.016 & 2.9760 & 2.6776 & 2.6776  \\  
SDSSJ0145-0945  & C            &  2.732 & 2.6948 & 2.5000 & 0.0000  \\  
SDSSJ0145-0945B & C            &  2.731 & 2.6939 & 2.5000 & 0.0000  \\  
SDSSJ0148-0907  & C            &  3.322 & 3.2790 & 2.9949 & 2.9949  \\  
SDSSJ0209-0005  & C-(1)        &  2.858 & 2.8196 & 2.5230 & 2.5230  \\  
SDSSJ0251-0737  & C            &  3.084 & 3.0433 & 2.5000 & 0.0000  \\  
SDSSJ0303-0023  & C-(1)        &  3.229 & 3.1869 & 2.9405 & 2.9405  \\  
SDSSJ0304-0008  & C-(1)        &  3.297 & 3.2542 & 2.5354 & 2.5354  \\  
SDSSJ0338-0005  & C            &  3.053 & 3.0126 & 2.7460 & 2.7460  \\  
SDSSJ0859+0205  & C            &  2.979 & 2.9394 & 2.8455 & 2.8455  \\  
SDSSJ0908+0658  & C            &  3.062 & 3.0216 & 2.5000 & 0.0000  \\  
SDSSJ0915+0549  & C            &  2.976 & 2.9364 & 2.6632 & 2.6632  \\  
SDSSJ0931-0000  & C            &  3.209 & 3.1671 & 2.9265 & 2.9265  \\  
SDSSJ0937+0417  & C            &  3.127 & 3.0859 & 2.5000 & 0.0000  \\  
SDSSJ0942+0422  & C            &  3.283 & 3.2404 & 2.4706 & 2.4706  \\  
SDSSJ0947+1421  & C            &  3.041 & 3.0008 & 2.5000 & 0.0000  \\  
SDSSJ1004+0018  & C            &  3.050 & 3.0097 & 2.7460 & 2.7460  \\  
SDSSJ1015+1118  & C            &  2.913 & 2.8740 & 2.8700 & 2.8700  \\  
SDSSJ1019+0825  & C            &  2.989 & 2.9493 & 2.9493 & 2.9647  \\  
SDSSJ1025+0452  & C            &  3.243 & 3.2008 & 3.1296 & 3.1296  \\  
SDSSJ1205-0048  & C            &  2.986 & 2.9463 & 2.5000 & 0.0000  \\  
SDSSJ1405+0507  & C-B          &  2.994 & 2.9542 & 2.5000 & 0.0000  \\  
SDSSJ1432+1139  & C-B          &  3.006 & 2.9661 & 2.5000 & 0.0000  \\  
SDSSJ1459+0024  & C            &  3.024 & 2.9839 & 2.7675 & 2.7675  \\  
SDSSJ1513+0855  & C            &  2.891 & 2.8523 & 2.5000 & 0.0000  \\  
SDSSJ1521-0048  & C            &  2.949 & 2.9097 & 2.5000 & 0.0000  \\  
SDSSJ1550+0537  & C            &  3.153 & 3.1117 & 2.9798 & 2.9798  \\  
SDSSJ1551+0908  & C            &  2.755 & 2.7176 & 2.6997 & 2.6997  \\  
SDSSJ1608+0715  & C            &  2.888 & 2.8493 & 2.5000 & 0.0000  \\  
SDSSJ1615+0608  & C-B          &  3.062 & 3.0216 & 2.9882 & 2.9882  \\  
SDSSJ1620+0941  & C            &  3.061 & 3.0206 & 2.8236 & 2.8236  \\  
SDSSJ2038-0025  & C            &  2.707 & 2.6701 & 2.4896 & 2.4896  \\  
SDSSJ2049-0554  & C            &  3.195 & 3.1532 & 2.5000 & 0.0000  \\  
SDSSJ2055-0511  & C-B          &  3.004 & 2.9641 & 2.5000 & 0.0000  \\  
SDSSJ2100-0641  & C            &  3.130 & 3.0889 & 3.0889 & 3.0913  \\  
SDSSJ2222-0946  & C            &  2.914 & 2.8750 & 2.5000 & 0.0000  \\  
SDSSJ2225-0041  & C            &  2.774 & 2.7364 & 2.5575 & 2.5575  \\  
SDSSJ2238-0921  & C            &  3.278 & 3.2354 & 3.1280 & 3.1280  \\  
SDSSJ2319-1040  & C            &  3.180 & 3.1384 & 2.6749 & 2.6749  \\  
SDSSJ2334-0908  & C            &  3.351 & 3.3077 & 3.2260 & 3.2260  \\  
SDSSJ2348-1041  & C            &  3.153 & 3.1117 & 2.9975 & 2.9975    
\enddata                                          	    	  					       
\tablenotetext{a}{C: color selected; NC: non color selected; B: broad absorption line quasar (visually classified); (1): 
included in the survey by \citet{sar89}; (2): included in the survey by \citet{lan91}.} 
\end{deluxetable*}

%% file: table3.tex
\begin{deluxetable*}{lccccccccc}
\tablewidth{0pc}
\tablecaption{Analysis of the metal lines in the LLS composite spectrum.\label{tabew}}
\tablehead{
\colhead{Ion}
& \colhead{$\lambda_{\rm rest}$} 
& \colhead{$f$} 
& \colhead{$W$\tablenotemark{a}} 
& \colhead{$\sigma_{\rm W}$} 
& \colhead{sign}
& \colhead{$\log N_{\rm X,lin}$\tablenotemark{a,c}}
& \colhead{$\log N_{\rm X,aodm}$\tablenotemark{a}}
& \colhead{$\sigma_{\rm N,aodm}$}
& \colhead{Adopted Value}\\
& (\AA) 
&  
& (\AA) 
& (\AA) 
&
& (cm$^{-2})$
& (cm$^{-2})$
& (cm$^{-2})$
& }
\startdata 
     AlII    		    &	1670.7874  & 1.8800  &  0.039 &  0.010 & $=$ & 11.93&   11.94& 0.12  & $11.94\pm0.12$ \\
     AlIII   		    &	1854.7164  & 0.5390  &  0.096 & -      & $<$ & 12.77&   12.78&   -   & $<12.78$  \\
     AlIII   		    &	1862.7895  & 0.2680  &  0.076 & -      & $<$ & 12.97&   12.97&   -   & \\
     CII     		    &	1334.5323  & 0.1278  &  0.063 &  0.009 & $=$ & 13.49&   13.51& 0.06  & $13.51\pm 0.06$ \\
     CIII\tablenotemark{b}  &    977.0200  & 0.7620  &  0.655 &  0.067 & $>$ & 14.01&   14.24& 0.09  & $>14.24$ \\
     CIV     		    &	1548.1950  & 0.1908  &  0.218 &  0.010 & $=$ & 13.73&   13.79& 0.02  & $13.81\pm  0.04$ \\
     CIV     		    &	1550.7700  & 0.0952  &  0.127 &  0.009 & $=$ & 13.80&   13.82& 0.03  &  \\
     FeII    		    &	1608.4511  & 0.0580  &  0.047 & -      & $<$ & 13.55&   13.56&   -   & $<13.56$ \\
     OI      		    &	1302.1685  & 0.0489  &  0.121 & -      & $<$ & 14.22&   14.24&   -   & $<14.24$  \\
     SiII    		    &	1304.3702  & 0.0940  &  0.116 & -      & $<$ & 13.91&   13.93&   -   & $<13.29$  \\
     SiII    		    &	1526.7066  & 0.1270  &  0.050 & -      & $<$ & 13.28&   13.29&   -   &   \\
     SiII    		    &	1808.0130  & 0.0022  &  0.084 & -      & $<$ & 15.12&   15.13&   -   &   \\
     SiIII\tablenotemark{b} &   1206.5000  & 1.6600  &  0.524 &  0.085 & $>$ & 13.39&   13.57& 0.10  & $>13.57$\\
     SiIV    		    &	1393.7550  & 0.5280  &  0.156 &  0.008 & $=$ & 13.24&   13.27& 0.02  & $13.31 \pm 0.05$ \\
     SiIV    		    &	1402.7700  & 0.2620  &  0.095 &  0.010 & $=$ & 13.32&   13.34& 0.05  &   \\
     ZnII    		    &	2026.1360  & 0.4890  &  0.077 & -      & $<$ & 12.63&   12.64&   -   & $<12.64$
\enddata 
\tablenotetext{a}{For undetected transitions, we list the $3\sigma$ upper limits.}              
\tablenotetext{b}{At MagE spectral resolution, this transition is likely to be saturated.}  
\tablenotetext{c}{The listed column densities are subject to significant uncertainties as discussed in the text.}  
\end{deluxetable*}